\documentclass[AMA,STIX1COL]{WileyNJD-v2}

\articletype{Survey}%

\received{26 April 2016}
\revised{6 June 2016}
\accepted{6 June 2016}

\usepackage{balance,listings}
\usepackage{amsmath}
\usepackage{tcolorbox}

\usepackage{subfigure,booktabs}
\usepackage{multirow,threeparttable,makecell,color,colortbl,graphicx}
\usepackage{indentfirst}
\usepackage{supertabular,longtable}
\usepackage{rotating}
\usepackage{amssymb} % for checkmark
\usepackage{url}

\def\BibTeX{{\rm B\kern-.05em{\sc i\kern-.025em b}\kern-.08em
    T\kern-.1667em\lower.7ex\hbox{E}\kern-.125emX}}

\graphicspath{{pics/}}

\begin{document}

\title{The Progress, Challenges, and Perspectives of Directed Greybox Fuzzing}

\author[1]{Pengfei Wang}

\author[1]{Xu Zhou}

\author[1]{Tai Yue}

\author[1]{Peihong Lin}

\author[1]{Yingying Liu}

\author[1]{Kai Lu}

\authormark{Pengfei Wang \textsc{et al}}

\address[1]{\orgdiv{College of Computer}, \orgname{National University of Defense Technology}, \orgaddress{\state{Changsha}, \country{China}}}

%\address[2]{\orgdiv{Org Division}, \orgname{Org Name}, \orgaddress{\state{State name}, \country{Country name}}}

%\address[3]{\orgdiv{Org Division}, \orgname{Org Name}, \orgaddress{\state{State name}, \country{Country name}}}

\corres{*Corresponding author: Pengfei Wang. \email{pfwang@nudt.edu.cn}}

\presentaddress{No.109 Deya Road, Kaifu District, Changsha, China}

\abstract[Summary]{Greybox fuzzing is a scalable and practical approach for software testing. 
Most greybox fuzzing tools are coverage-guided as reaching high code coverage is more likely to find bugs. However, since most covered codes may not contain bugs, blindly extending code coverage is less efficient, especially for corner cases. 
Unlike coverage-guided greybox fuzzing which increases code coverage in an undirected manner, directed greybox fuzzing (DGF) spends most of its time allocation on reaching specific targets (e.g., the bug-prone zone) without wasting resources stressing unrelated parts. Thus, DGF is particularly suitable for scenarios such as patch testing, bug reproduction, and special bug detection. For now, DGF has become an active research area.
However, DGF has general limitations and challenges that are worth further studying.
Based on the investigation of 42 state-of-the-art fuzzers that are closely related to DGF, we conduct the first in-depth study to summarize the empirical evidence on the research progress of DGF.
This paper studies DGF from a broader view, which takes into account not only the location-directed type that targets specific code parts, but also the behavior-directed type that aims to expose abnormal program behaviors. 
By analyzing the benefits and limitations of DGF research, we try to identify gaps in current research, meanwhile, reveal new research opportunities, and suggest areas for further investigation.}

\keywords{directed greybox fuzzing, location directed fuzzing, target directed fuzzing}

%\jnlcitation{\cname{%
%\author{Pengfei W},
%\author{B. Hoskins},
%\author{R. Lee},
%\author{G. Masato}, and
%\author{T. Woollings}} (\cyear{2016}),
%\ctitle{A regime analysis of Atlantic winter jet variability applied to evaluate HadGEM3-GC2}, \cjournal{Q.J.R. Meteorol. Soc.}, \cvol{2017;00:1--6}.}

\maketitle

%\footnotetext{\textbf{Abbreviations:} ANA, anti-nuclear antibodies; APC, antigen-presenting cells; IRF, interferon regulatory factor}

\section{Introduction}
\label{intro}

Fuzzing is an automated software testing approach proposed by Barton Miller in 1989 \cite{miller1990empirical}. 
By repeatedly and randomly mutating the inputs to the program under test (PUT), fuzzing is effective and practical in vulnerability detection.
As one of the most efficient and scalable fuzzing categories, greybox fuzzing \cite{manes2019art, godefroid2020fuzzing, boehme2021fuzzing, zhu2022fuzzing, liang2018fuzzing, li2018fuzzing} develops rapidly in recent years.
Based on the feedback information from the execution of the PUT, greybox fuzzers use an evolutionary algorithm to generate new inputs and explore paths. Greybox fuzzing is widely used to test application software, libraries \cite{blair2020hotfuzz, joffe2019directing}, kernel code \cite{schumilo2017kafl, song2019periscope, kimhfl}, and  protocols \cite{yu2019poster, zheng2019firm, garbelini2020greyhound}. %, smart contracts \cite{jiang2018contractfuzzer, wustholz2019targeted}, and multi-threaded programs \cite{zhaokrace, vinesh2020confuzz, jeong2019razzer}.
Most greybox fuzzing tools are coverage-guided, which aim to cover as many program paths as possible within a limited time budget. This is because, intuitively, reaching high code coverage is more likely to find bugs. However, it is not appropriate to treat all codes of the program as equal because most covered codes may not contain bugs. For example, according to Shin \textit{et al.}~\cite{shin2013can}, only 3\% of the source code files in Mozilla Firefox have vulnerabilities. Thus, testing software by blindly extending code coverage is less efficient, especially for corner cases. Since achieving full code coverage is difficult in practice, researchers have been trying to target the vulnerable parts in a program to improve efficiency and save resources. Directed fuzzing is proposed as a means of achieving this aim \cite {ganesh2009taint}.

Unlike coverage-based fuzzers that blindly increase the path coverage, a directed fuzzer focuses on target locations (e.g., the bug-prone zone) and spends most of its time budget on reaching these locations without wasting resources stressing unrelated parts. Originally, directed fuzzers were based on symbolic execution \cite{ganesh2009taint, ma2011directed, person2011directed, marinescu2013katch}, which uses program analysis and constraint solving to generate inputs that exercise different program paths. Such directed fuzzers cast the reachability problem as an iterative constraint satisfaction problem~ \cite {bohme2017directed}. However, since directed symbolic execution (DSE) relies on heavy-weight program analysis and constraint solving, it suffers from scalability and compatibility limitations.

In 2017, B\"ohme \textit{et al.} introduced the concept of directed greybox fuzzing (DGF) \cite{bohme2017directed}. Greybox fuzzing generates inputs by mutating seeds. By specifying a set of target sites in the PUT and leveraging lightweight compile-time instrumentation, a directed greybox fuzzer can use the distance between the input and the target as the fitness function to assist seed selection. By giving more mutation chances to seeds that are closer to the target, it can steer the greybox fuzzing to reach the target locations gradually. Unlike traditional fuzzing techniques, DGF casts reachability as an optimization problem whose aim is to minimize the distance between generated seeds and their targets \cite {bohme2017directed}. Compared with directed symbolic execution, DGF has much better scalability and improves efficiency by several orders of magnitude. For example, B\"ohme \textit{et al}. can reproduce the Heartbleed vulnerability within 20 minutes, while the directed symbolic execution tool KATCH \cite{marinescu2013katch} needs more than 24 hours \cite{bohme2017directed}. 
%Automated target recognition is a key feature of current DGF tools, and thus they are capable of both identifying their targets and honing in upon their response. 

\textbf{Motivation}. 
For now, DGF has become a research hot spot and it is growing very fast. It has evolved beyond the original pattern that depends on manually labeled target sites and distance-based metrics to prioritize the seeds. 
New fitness metrics, such as trace similarity and vulnerability prediction models, are used. 
Current DGF tools can not only identify targets automatically but also expose target program behavior in a directed manner.
A great number of variations have been used to boost software testing under different scenarios, such as patch testing \cite{you2017semfuzz, peng20191dvul, nguyen2020binary}, regression testing \cite{zhangdeltafuzz,zhu2021regression}, 
bug reproduction \cite{wangnot, kim2019poster, nguyen2020binary}, knowledge integration \cite{aschermannijon}, result validation \cite{li2019v, zhao2019suzzer, zhu2020defuzz, osterlundparmesan}, energy-saving \cite{garbelini2020greyhound}, and special bug detection \cite{wen2020memlock, wang2020uafl, nguyen2020binary, kim2019rvfuzzer, lemieux2018perffuzz, petsios2017slowfuzz, garbelini2020greyhound}. 
Though fast-growing and useful, DGF has general limitations and challenges that are worth further study.
Under this background, we conduct this work to summarize the empirical evidence on the research progress of DGF.
Based on the analysis of benefits and limitations of DGF research, we try to identify gaps in current research, meanwhile, reveal new research opportunities, and suggest areas for further investigation.

\textbf{Research Questions}.
We conduct the first in-depth study of DGF in this work.
To study DGF from a broader view, we take into account not only the location-directed type that targets specific code parts, but also the behavior-directed type that targets exposing abnormal program behaviors to find bugs. 
In summary, we design the following research questions:

\begin{itemize}

\item  \textbf{RQ1}: How the target identification method is changed in the up-to-date research of DGF?

\item  \textbf{RQ2}: In addition to distance, are there any new fitness metrics in the recent development of DGF?

\item  \textbf{RQ3}: How the recent DGF tools are optimized regarding the key steps of fuzzing? 

\item  \textbf{RQ4}: What are the challenges of the DGF research?  Are there any potential solutions?

\item \textbf{RQ5}: What are the typical application of DGF? How to choose a DGF tool for a specific application scenario?

\item  \textbf{RQ6}: What are the perspectives of the future trends on DGF research?

\end{itemize}

In this work, we make the following contributions. 
\begin{itemize}

\item [-] We investigate 42 state-of-the-art fuzzers that are closely related to DGF to systemize recent progress in the field and answer research questions \textbf{RQ1}, \textbf{RQ2}, \textbf{RQ3};

\item [-] Based on the analysis of the known works, a summary of five challenges to DGF research is provided. We discuss these challenges with examples and disclose the deep reasons behind them, aiming to propose possible solutions to address them and answer \textbf{RQ4};
%, including differentiated weight metric, global optimum discrepancy, exploration-exploitation coordination, multi-objective optimization, and binary code support. %We disclose the deep reasons behind these challenges and propose possible solutions to address them.

\item [-] Based on the fast-growing rate of DGF tools, we summarize the typical application scenarios of DGF and provide suggestions on how to choose a DGF tool for a specific application scenario, which answers \textbf{RQ5};

\item [-] We make suggestions in terms of the perspectives for the research points of DGF that are worth exploring in the future, aiming to facilitate and boost research in this field and answer \textbf{RQ6}.

\end{itemize}

%The rest of the paper is organized as follows: Section 2 reviews the background knowledge of directed testing approaches. Section 3 evaluates state-of-the-art directed greybox fuzzers based on the extracted metrics and systemizes the techniques used in each core component of DGF. Section 4 summarizes the challenges of this field based on current research progress. Section~5 discusses the application scenarios of DGF. Section 6 discusses future perspectives and is followed by conclusions.

\section{Background}
\label{background}
%This section provides background knowledge of CGF and DGF. We use AFL and AFLGo to illustrate the principle, respectively. We also introduce the related works of directed symbolic execution and search-based software testing. Then we compare DGF, CGF, DSE, and SBST to show the difference. Finally, we summarize the application scenarios of DGF.
%Most of the state-of-the-art directed greybox (and hybrid) fuzzers are built on top of AFL. We introduce the critical techniques of AFL to facilitate the understanding of DGF.

\subsection{Blackbox Fuzzing}
Blackbox fuzzing is the original and simplest form of fuzzing \cite{miller1990empirical}.
The workflow of blackbox fuzzing is represented by the blue long dashed lines in Figure \ref{cgf}.
It randomly mutates the inputs and then tests the PUT with these modified inputs.
The procedure can be repeatedly executed to generate as many new inputs as needed. 
However, since blackbox fuzzing does not support any feedback scheme, the inputs are generated blindly.
Thus, the code coverage is usually low, and most of the generated inputs are invalid or redundant, resulting in low testing efficiency.
Though simple and less efficient, blackbox fuzzing can be effective in finding security vulnerabilities, especially 
for the testing scenarios where the feedback scheme of greybox fuzzing is difficult to realize,
such as protocol testing \cite{feng2021snipuzz, gascon2015pulsar, yu2019poster,shu2022iotinfer,de2015protocol}.

\subsection{Coverage-guided Greybox Fuzzing}

\begin{figure}
\centering
\includegraphics[width=0.8\columnwidth]{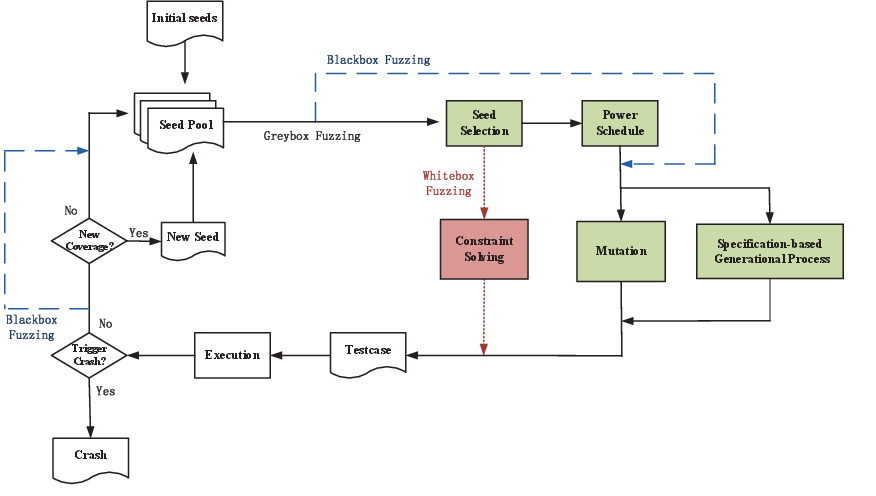}
\caption{Workflow of different fuzzing approaches.}
\label{cgf}
\end{figure}

Coverage-guided greybox fuzzing (CGF) aims to maximize the code coverage to find hidden bugs. 
By introducing a feedback scheme, the fuzzing status (i.e., whether a new path is explored) can be leveraged to guide the generation of new inputs.
%A proper seed is selected from the inputs and mutated with a reasonable strategy. As a result, the new input generated is more likely to find a new path.  
Figure \ref{cgf} shows the workflow of CGF, where the components shown in the white boxes are the common steps used by almost all the fuzzing approaches, while the components shown in the green boxes are critical steps for CGF.
The testcase in CGF can be generated in a mutational or generational way, and a typical mutational CGF includes the following steps:

\begin{itemize}
\item \textbf{Seed Selection}. Select a proper seed (usually considered as high quality) from the seed queue for mutation;

\item \textbf{Power Schedule}. Determine the energy (i.e., the number of mutation chances) assigned to the selected seed;

\item \textbf{Mutation}. Mutate the seed with a set of pre-defined mutation strategies to generate test inputs;

\item \textbf{Execution}. Run the PUT with the test inputs to monitor whether a new path is exercised. If a new path is triggered, the input is added to the queue as a new seed.

\end{itemize}

Here we take the widely used tool AFL (American fuzzy lop) \cite{afl} as a representative to illustrate the principle of CGF. 
AFL is a prevalently used coverage-based greybox fuzzer, and many state-of-the-art greybox fuzzers \cite{bohme2017coverage, lyu2019mopt, chen2019matryoshka, yue2019learnafl, Zhang2022MobFuzz} are built on top of it. 
AFL uses lightweight instrumentation to capture basic block transitions at compile-time and gain coverage information during run-time. It then selects a seed from the seed queue and mutates the seed to generate test cases. If a test case exercises a new path, it is added to the queue as a new seed. AFL favors seeds that trigger new paths and gives them preference (i.e., more energy) over the non-favored ones. Compared to other instrumented fuzzers, AFL has a modest performance overhead. However, though some tools (e.g., Fairfuzz \cite{lemieux2017fairfuzz}) try to reach the rare part of a code, most greybox fuzzers treat all codes of the program as equal. Thus, CGF is less efficient as effort is wasted on non-buggy areas.

\subsection{Directed Greybox Fuzzing}

\begin{table}[t]
 
\vspace{2ex}
 
\centering
\begin{tabular}{p{1.2cm}@{}p{10cm}}
\hline
 
&Algorithm 1: Directed greybox fuzzing scheme. \\   
\hline
\textbf{Input}: &\textit{i} $-$ Initial input \\
\textbf{Input}: &\textit{Target} $-$ A set of target locations \\
\textbf{Output}: &\textit{BugInput} $-$ A set of buggy input \\ 

01 & $BugInput \leftarrow \varnothing$ \\
%01 & $Targets \leftarrow {t, ... ,t}$ \\
02 & $SeedQueue \leftarrow i$ \\ 
03 & \textbf{while} true \textbf{do} \\
04 & \quad $ s \leftarrow select(SeedQueue)$\\
05 & \quad $ s' \leftarrow mutation(s)$\\
06 & \quad $ trace \leftarrow execution(s')$ \\
07 & \quad \textbf{if} $find\_new\_path(trace) $ \textbf{then} \\
08 & \quad\quad $ SeedQueue \leftarrow SeedQueue + s'$ \\ 
09 & \quad \textbf{if} $trigger\_crash(trace)$ \textbf{then} \\
10 & \quad\quad $ BugInput \leftarrow BugInput + s'$ \\
11 & \quad $ distance \leftarrow evaluate\_seed(trace, Targets)$ \\
12 & \quad $ SeedQueue \leftarrow sort\_insert(SeedQueue, s', distance)$ \\
13 & \textbf{end} \\

\hline
\end{tabular}
\label{algorithm1}
\vspace{-3ex}
\end{table}

%In 2017, B\"ohme \textit{et al}. introduced the concept of DGF and implemented a tool called AFLGo~\cite{bohme2017directed} based on the modern greybox fuzzing framework.Unlike CGF blindly increasing the path coverage, DGF aims to reach a set of pre-identified locations in the code (potentially the buggy parts) and spends most of its time budget on reaching target locations without wasting resources stressing unrelated parts. 

Unlike CGF which blindly increases path coverage, DGF aims to reach a set of pre-identified locations in the code (potentially the buggy parts) and spends most of its time budget on reaching target locations without wasting resources stressing unrelated parts. 
To describe the DGF principle, we use AFLGo \cite{bohme2017directed} as an example. AFLGo follows the same general principles and architecture as CGF. At compile-time, except for instrumentation, AFLGo also calculates the distances between the input and pre-defined targets. The distance is calculated as the average weight of the execution trace to the target basic blocks. 
The execution trace weight is determined by the number of edges in the call graph and control-flow graphs of the program. Then, at run-time, AFLGo \cite{bohme2017directed} prioritizes seeds based on distance instead of new path coverage and gives preference to seeds closer to the targets at basic block level distance. B\"ohme \textit{et al}. \cite{bohme2017directed} view the greybox fuzzing process as a Markov chain that can be efficiently navigated using a ``power schedule''. They leveraged a simulated annealing strategy to gradually assign more energy to a seed that is closer to targets than to a seed that is further away. They cast reachability as an optimization problem to minimize the distance between the generated seeds and their targets~\cite{bohme2017coverage}. 
Similar to CGF, in Figure \ref{cgf}, the components shown in the green boxes are also critical steps for DGF, and DGF is mainly optimized via these steps.

\textbf{The exploration-exploitation problem}.
DGF fuzzing is a two-part method, which is readily separated into phases of \textit{exploration} and \textit{exploitation} \cite{bohme2017directed}. The exploration phase is designed to uncover as many paths as possible. Like many coverage-guided fuzzers, DGF in this phase favors seeds that trigger new paths and prioritizes them. This is because new paths increase the potential to lead to targets, and it is particularly necessary when the initial seeds are quite far from their targets. Then, based on the known paths, the exploitation phase is invoked to drive the engine to the target code areas. In this phase, B\"ohme \textit{et al}. \cite{bohme2017directed} prioritize seeds that are closer to the targets and assign more energy to them. The intuition behind this is that if the path that the current seed executes is closer to any of the expected paths that can reach the target, more mutations on that seed should be more likely to generate expected seeds that fulfill the demands. The exploration-exploitation trade-off lies in how to coordinate these two phases. B\"ohme \textit{et al}. \cite{bohme2017directed} use a fixed splitting of the exploration and exploitation phases. For example, in a 24-hour test, AFLGo assigns 20 hours for exploration and then 4 hours for exploitation.

\subsection{Directed Whitebox Fuzzing}
A directed whitebox fuzzer \cite{godefroid2008automated} is mostly implemented into a symbolic execution engine such as KLEE \cite{cadar2008klee}, KATCH \cite{marinescu2013katch}, and BugRedux \cite{jin2012bugredux}.
Directed Symbolic Execution (DSE) uses program analysis and constraint solving to generate inputs that systematically and effectively explore the state space of feasible paths \cite{ma2011directed}. In Figure \ref{cgf}, the constraint solving component in the red box is mandatory for whitebox fuzzing but optional for DGF.
Once a target path is identified, potential solutions to the path constraints are explored by creating test cases. Since most paths are actually infeasible, the search usually proceeds iteratively by finding feasible paths to intermediate targets. Unlike DGF, which casts reachability as an optimization problem to minimize the distance between generated seeds and their targets \cite{bohme2017directed}, DSE casts the reachability problem as an iterative constraint satisfaction problem \cite{bohme2017directed}. DSE is effective in various scenarios, such as reaching error-prone program locations (e.g., critical syscalls \cite{haller2013dowsing}), testing code patches \cite{bohme2013partition, marinescu2013katch, santelices2008test}, exercising corner paths to increase coverage \cite{xu2010directed}, and reproducing failures in-house \cite{jin2012bugredux, rossler2013reconstructing}.

However, DSE’s effectiveness comes at the cost of efficiency. 
The heavy-weight program analysis and constraint solving of DSE is rather time-consuming. 
At each iteration, DSE utilize program analysis techniques to identify branches that can be negated to get closer to the target. Then, based on the sequence of instructions along these paths, it constructs the corresponding path conditions. Finally, it checks the satisfiability of those conditions using a constraint solver. 
DGF is capable of producing a far greater number of inputs in a given timeframe than DSE can achieve \cite{bohme2017directed}. B\"ohme \textit{et al.} have demonstrated with experiments that DGF outperforms DSE both in terms of effectiveness and efficiency. For example, AFLGo can expose the Heartbleed vulnerability in 20 minutes while the DSE tool KATCH cannot even in 24 hours ~\cite{bohme2017directed}.

%Directer~\cite{song2018directer},

\subsection{Search-based Software Testing}
Search-based Software Testing (SBST) formulates a software testing problem into a computational search problem that can be optimized with meta-heuristic search techniques, such as hill-climbing, simulated annealing, and genetic algorithms \cite{mcminn2004search}. The key to the optimization process is defining a problem-specific fitness function, which guides the search by measuring the quality of potential solutions from a possibly infinite search space. Greater fitness values are assigned to those inputs that provide data closer to the focal point in the program \cite{mcminn2011search}. The original use of SBST was structural coverage testing \cite{miller1976automatic}, including path and branch coverage. The path taken through the program under test is compared with some structure of interest for which coverage is sought \cite {mcminn2011search}. The fitness function usually incorporates two metrics—approach level and branch distance \cite{wegener2001evolutionary}. The complete fitness value is computed by normalizing the branch distance and adding it to the approach level \cite{mcminn2011search}. 
In addition to structural testing, SBST can also be used for functional testing \cite{buehler2003evolutionary, buhler2008evolutionary}, temporal testing \cite{puschner1998testing, wegener1997testing, wegener1998verifying}, robustness testing \cite{schultz1993test}, integration testing \cite{briand2002using, colanzi2011integration}, regression testing \cite{li2007search}, stress testing \cite{briand2005stress}, mutation testing \cite{jia2008constructing}, interaction testing\cite{cohen2003constructing, petke2013efficiency, cohen2007interaction}, state testing\cite{derderian2006automated, derderian2006, lehre2014runtime}, and exception testing \cite{tracey2000automated, tracey2002search}. The main difference between fuzzing and SBST is that fuzzing uses lightweight and scalable heuristics to maintain and evolve a population of test inputs, whereas SBST approaches typically use an optimization formula to search for ideal test cases~\cite{medicherla2020fitness}.

Though SBST and DGF are close, they are different. For example, fuzzing uses lightweight instrumentation to guide the evolution, thus, it is simpler than the computational search optimization in SBST. Besides, the search optimization of SBST might be trapped by the local optimal solution, while DGF can escape the local optimal solution more quickly due to the randomness of fuzzing. %Thus, we think DGF is more efficient and practical than SBST. Since SBST is a relatively mature direction, 
In this paper, we focus on DGF and provide the most up-to-date research progress.

\section{Methodology}
This section introduces the methodology we adopted when conducting this research. The motivation and research questions have been introduced in Section \ref{intro}, thus, here we only describe the other key elements in a review protocol.

\subsection{Inclusion and Exclusion Criteria}
\label{criteria}
This paper defines a tool as a directed greybox fuzzer from a broader view, namely, that a fuzzer either reaches specific target locations or triggers specific program buggy behavior by optimizing a customized fitness function. The following inclusion criteria are thus specified, which also serve as the definition of DGF in this paper.

\begin{itemize}

\item [-] The core mechanism should be greybox fuzzing, which relies on the instrumentation of the PUT and includes the key steps of seed prioritization, power scheduling, and mutator scheduling;%Thus, we exclude directed white-box fuzzers based on symbolic execution (still discuss the relationship in Section \ref{background}). However, we include the mechanism of hybrid fuzzing to emphasize the importance of symbolic execution.

\item [-]  The directedness is realized by optimizing the fitness metric in the key steps of greybox fuzzing, which includes input optimization, seed prioritization, power scheduling, mutator scheduling, and mutation operations;

\item [-] The fitness goal is to reach a specific site or to trigger certain buggy behavior of a program. The site could be a manually labeled target or a potential bug location predicted automatically, such as by machine learning \cite{li2019v, zhao2019suzzer, zhu2020defuzz} or static analysis \cite{osterlundparmesan}. The target buggy behavior could be a non-functional property (e.g., memory consumption ~\cite{wen2020memlock}), or a certain bug type (e.g., algorithmic complexity vulnerability~\cite{lemieux2018perffuzz}).  

\end{itemize}

We classify a DGF tool as directed for target location type when its object is reaching target sites, and the fitness metric can be measured visibly on a concrete structure, such as on the execution trace, the control-flow graph, or call-graph.
The target can be a single location, a set of basic blocks, or a sequence of ordered call sites. In contrast, if a DGF tool is directed with a certain fitness metric but without a fixed target, then it is classified as directed for target behavior. For this type, the targets need not or cannot be pre-labeled, and the fitness metric is not as visible as the first type. With the optimization of the fitness function, a target can be reached automatically and the buggy behavior will be exposed.

However, to concentrate on the research of DGF, the following types of papers will be excluded.

\begin{itemize}
\item  [-] Directed whitebox fuzzing realized only via symbolic execution (we still include directed hybrid fuzzing papers that assist DGF with symbolic execution), 

\item [-] Papers on search-based software testing,

\item  [-] Informal literature reviews and technical reports,

\item  [-] Too short papers (less than 4 pages) without a clear description of the approach or the evaluation.
\end{itemize}

\subsection{Search Process}

The search process consists of three rounds. The first round is a manual search of specific conference proceedings and journal papers via an academic search engine with keywords, which includes the following steps.

(1) The publications are initially collected from the proceedings of the top-level conferences on security and software engineering since 2017. Alphabetically, ACM Conference on Computer and Communications Security (CCS), IEEE Symposium on Security and Privacy (S\&P), USENIX Security Symposium (Sec), Network and Distributed System Security Symposium (NDSS), and International Conference on Software Engineering (ICSE). ACM International Symposium on the Foundations of Software Engineering ESEC/FSE), IEEE/ACM International Conference on Automated Software Engineering (ASE). We search with ``directed greybox fuzzing'' and ``directed fuzzing''. We collected 19 papers.

(2) Then, we use google scholar to search for works from journals and preprints by searching with keywords including  ``directed greybox fuzzing'', ``directed fuzzing'', `` targeted fuzzing''. We collected 15 papers.

(3) After that, we refer to a popular fuzzing paper repository \footnote{Recent Fuzzing Papers, https://wcventure.github.io/FuzzingPaper/\#survey-of-directed-fuzzy-technology} and manually select papers related to DGF. We also refer to another paper repository that only collects papers related to directed fuzzing \footnote{Awesome directed fuzzing, \url{https://github.com/uafuzz/awesome-directed-fuzzing}}. We collected 21 papers.

Then, in the second round, we filter out the duplicates from the collection in the previous round. When a paper has been published in more than one journal/conference, the most complete version will be used. As a result, 49 papers remained.

In the third round, we read each paper we collected and filter out the papers based on the research content with the inclusion and exclusion criteria from Section \ref{criteria}. Finally, 42 papers ranging from 2017.1 to 2022.5 (listed in Table 1) remained for further investigation. 

\subsection{Data collection}
First, we interview at least ten researchers to list the important aspects of DGF tools they care about. The researcher includes postgraduate students, PhD students, and faculties. Then, we summarized the suggestions and extracted the aspects that they care about most. Based on this, the data extracted from each paper will be:

\begin{itemize}
\item The publication source (i.e. the conference, journal, or preprint) and year.

\item The fitness goal. To reach what kind of target sites (e.g., vulnerable function) or to expose what target bugs?

\item The fitness metric used in the evolutionary process of fuzzing. For example, the distance to the targets. 

\item How the targets are identified or labeled? For example, predicted by deep learning models.

\item The implementation information. What tool is the fuzzer implemented based on? Is the fuzzer open-sourced?

\item Whether the tool supports binary code analysis?

\item Whether the tool supports kernel analysis?

\item Whether the tool supports multi-targets searching?

\item Whether the tool supports multi-objective optimization?

\item What key steps in fuzzing are optimized to realize the directedness? Namely input optimization, seed prioritization, power scheduling, mutator scheduling, and mutation operations.

\item What techniques are adopted in the optimization? Namely control-flow analysis, static analysis, data-flow analysis, machine learning, semantic analysis, and symbolic execution.

\end{itemize}

\subsection{Data Analysis}
The extracted data is tabulated (Table 1) to show the basic information about each study. Then we review the extracted data and try to answer the research questions as follows:

 \textbf{RQ1}: How the target identification method is changed in the up-to-date research on DGF? This will be addressed by summarizing how the targets of the documented research are identified or labeled. 

\textbf{RQ2}: In addition to distance, are there any new fitness metrics in the recent development of DGF? This will be addressed by summarizing the fitness metrics of the documented research. 

\textbf{RQ3}: How the recent DGF tools are optimized regarding the key steps of fuzzing? This will be addressed by analyzing the documented research on the optimization of the key steps of fuzzing (i.e., input optimization, seed prioritization, power scheduling, mutator scheduling, and mutation operations) to realize the directedness. 

\textbf{RQ4}: What are the challenges of the DGF research?  Are there any potential solutions? We will summarize comprehensive challenges for the DGF community based on the design and implementation of the documented research. For the design, we consider the fitness goal, fitness metric, how the targets are identified, and optimizations on the key fuzzing steps, while for implementation, we pay attention to efficiency and whether the tool supports binary, kernel, multi-targets, and multi-objectives.

\textbf{RQ5}: What are the typical application of DGF? How to choose a DGF tool for a specific application scenario? We will summarize the typical application of DGF based on the fitness goals, how the targets are identified, and the implementation details of the documented research.

\textbf{RQ6}: What are the perspectives and the future trends on DGF research? We will summarize future trends based on the analysis of the challenges and limitations of the current DGF research.

%\onecolumn
\begin{center}
\scriptsize
\renewcommand{\thetable}{1}
\label{tab:comparison}
\tablefirsthead{%
\hline
       \begin{sideways}Category \end{sideways}
        &\begin{sideways}Tools \end{sideways} 
  	    &\begin{sideways}Publication \end{sideways} 
        %&\begin{sideways}Fuzzing type \end{sideways}  
        &\begin{sideways}Fitness goal \end{sideways}
        &\begin{sideways}Fitness metric \end{sideways}      
        &\begin{sideways}Target identify \end{sideways}
        
        &\begin{sideways}Base tool \end{sideways}
        &\begin{sideways}Binary support \end{sideways}
        %&\begin{sideways}Indirect call support \end{sideways}
        &\begin{sideways}Kernel support \end{sideways}
        &\begin{sideways}Open sourced \end{sideways} 
        &\begin{sideways}Multi-targets \end{sideways}
        &\begin{sideways}Multi-objective \end{sideways} \\     
\hline
}
\tablehead{%
\hline
       \begin{sideways}Category \end{sideways}
        &\begin{sideways}Tools \end{sideways} 
  	    &\begin{sideways}Publication \end{sideways} 
        %&\begin{sideways}Fuzzing type \end{sideways}  
        &\begin{sideways}Fitness goal \end{sideways}
        &\begin{sideways}Fitness metric \end{sideways}      
        &\begin{sideways}Target identify \end{sideways}
        
        &\begin{sideways}Base tool \end{sideways}
        &\begin{sideways}Binary support \end{sideways}
        %&\begin{sideways}Indirect call support \end{sideways}
        &\begin{sideways}Kernel support \end{sideways}
        &\begin{sideways}Open sourced \end{sideways} 
        &\begin{sideways}Multi-targets \end{sideways}
        &\begin{sideways}Multi-objective \end{sideways} \\     
\hline
}
\tabletail{%
\hline
\multicolumn{12}{|r|}{\small\sl continued on next page}\\
\hline}
\tablelasttail{\hline}
\bottomcaption{This table is split across pages}
\tablefirsthead{%
\hline
 \begin{sideways}Category \end{sideways}
        &\begin{sideways}Tools \end{sideways} 
  	    &\begin{sideways}Publication \end{sideways} 
        %&\begin{sideways}Fuzzing type \end{sideways}  
        &\begin{sideways}Fitness goal \end{sideways}
        &\begin{sideways}Fitness metric \end{sideways}      
        &\begin{sideways}Target identify \end{sideways}
        
        &\begin{sideways}Base tool \end{sideways}
        &\begin{sideways}Binary support \end{sideways}
        %&\begin{sideways}Indirect call support \end{sideways}
        &\begin{sideways}Kernel support \end{sideways}
        &\begin{sideways}Open sourced \end{sideways} 
        &\begin{sideways}Multi-targets \end{sideways}
        &\begin{sideways}Multi-objective \end{sideways} \\  
\hline}
\tablehead{%
\hline
\multicolumn{12}{|l|}{\small\sl continued from previous page}\\
\hline
 \begin{sideways}Category \end{sideways}
        &\begin{sideways}Tools \end{sideways} 
  	    &\begin{sideways}Publication \end{sideways} 
        %&\begin{sideways}Fuzzing type \end{sideways}  
        &\begin{sideways}Fitness goal \end{sideways}
        &\begin{sideways}Fitness metric \end{sideways}      
        &\begin{sideways}Target identify \end{sideways}
        
        &\begin{sideways}Base tool \end{sideways}
        &\begin{sideways}Binary support \end{sideways}
        %&\begin{sideways}Indirect call support \end{sideways}
        &\begin{sideways}Kernel support \end{sideways}
        &\begin{sideways}Open sourced \end{sideways} 
        &\begin{sideways}Multi-targets \end{sideways}
        &\begin{sideways}Multi-objective \end{sideways} \\  
\hline}
\tabletail{%
\hline
\multicolumn{12}{|r|}{\small\sl continued on next page}\\
\hline}
\tablelasttail{\hline}
\bottomcaption{This table is split across pages}
\tablefirsthead{%
\hline
 \begin{sideways}Category \end{sideways}
        &\begin{sideways}Tools \end{sideways} 
  	    &\begin{sideways}Publication \end{sideways} 
        %&\begin{sideways}Fuzzing type \end{sideways}  
        &\begin{sideways}Fitness goal \end{sideways}
        &\begin{sideways}Fitness metric \end{sideways}      
        &\begin{sideways}Target identify \end{sideways}
        
        &\begin{sideways}Base tool \end{sideways}
        &\begin{sideways}Binary support \end{sideways}
        %&\begin{sideways}Indirect call support \end{sideways}
        &\begin{sideways}Kernel support \end{sideways}
        &\begin{sideways}Open sourced \end{sideways} 
        &\begin{sideways}Multi-targets \end{sideways}
        &\begin{sideways}Multi-objective \end{sideways} \\  
\hline}
\tablehead{%
\hline
\multicolumn{12}{|l|}{\small\sl continued from previous page}\\
\hline
 \begin{sideways}Category \end{sideways}
        &\begin{sideways}Tools \end{sideways} 
  	    &\begin{sideways}Publication \end{sideways} 
        %&\begin{sideways}Fuzzing type \end{sideways}  
        &\begin{sideways}Fitness goal \end{sideways}
        &\begin{sideways}Fitness metric \end{sideways}      
        &\begin{sideways}Target identify \end{sideways}
        
        &\begin{sideways}Base tool \end{sideways}
        &\begin{sideways}Binary support \end{sideways}
        %&\begin{sideways}Indirect call support \end{sideways}
        &\begin{sideways}Kernel support \end{sideways}
        &\begin{sideways}Open sourced \end{sideways} 
        &\begin{sideways}Multi-targets \end{sideways}
        &\begin{sideways}Multi-objective \end{sideways} \\  
\hline}
\tabletail{%
\hline
\multicolumn{12}{|r|}{\small\sl continued on next page}\\
\hline}
\tablelasttail{\hline}
\bottomcaption{Collection of directed greybox fuzzers}
\begin{supertabular}{p{0.4cm}|p{1.8cm}p{1.5cm}p{2cm}p{2.2cm}p{1.8cm}p{1.2cm}p{0.9cm}p{0.3cm}p{0.3cm}p{0.3cm}p{0.3cm}}

\hline
 
  \multirow{30}{0.4cm}{\begin{sideways} Directed for target location\end{sideways}} 
  \\                   
  
  &AFLGo \cite{bohme2017directed} &CCS'17
                                  %& G              %Fuzzing Type                                  
                                  & target sites   %fitness goal
                                  & distance       %fitness metric
                                  & manual label   %target identification
                                  
                                  & AFL           %base tool
                                  & $\times$      %Binary support
                                 % & $\times$      %Indirect call support
                                  & $\times$       %kernel support
                                  & $\checkmark$  %Open sourced
                                  & $\checkmark$  %Multi-targets support
                                  & $\times$       %Multi-object support                             
                                  \\
\hline

\end{supertabular}
\end{center}

%\twocolumn

\section{Research Progress on Directed Greybox Fuzzing}

\subsection{Overview}

Recently, DGF has been an active research area. To provide an overview of the DGF research, we summarize the following progress.

\begin{itemize}

\item In addition to the original fitness metric of distance, new fitness metrics have been adopted, such as sequence coverage, which is suitable for satisfying complex bug-triggering paths. Examples include UAFuzz \cite{nguyen2020binary}, UAFL \cite{wang2020uafl}, LOLLY \cite{liang2019sequence}, Berry \cite{liang2020sequence}, CAFL \cite{lee2021constraint}. Multi-dimensional fitness metrics are also proposed to detect hard to manifest vulnerabilities. Examples include AFL-HR \cite{medicherla2020fitness}, HDRFuzz \cite{medicherla2021hdr}, AFLPro \cite{ji2020aflpro};

\item To facilitate target identification, tools based on machine learning can predict and label potential targets automatically, examples include SUZZER \cite{zhao2019suzzer}, V-Fuzz \cite{li2019v}, DeFuzz \cite{zhu2020defuzz}, SemFuzz \cite{you2017semfuzz}. Meanwhile, CVE information, commit changes, binary diffing techniques, and tools such as UBSan and AddressSanitizer are adopted to label various potential vulnerable code regions. Examples include DrillerGo \cite{kim2019poster}, TortoiseFuzz \cite{wangnot}, AFLChurn \cite{zhu2021regression}, GREYHOUND \cite{garbelini2020greyhound}, DeltaFuzz \cite{zhangdeltafuzz}, 1DVUL \cite{peng20191dvul}, SAVIOR \cite{chen2019savior}, HDR-Fuzz \cite{medicherla2021hdr};

\item The fuzzing process has been enhanced with various approaches, such as using data-flow analysis and semantic analysis to generate valid input, using symbolic execution to pass complex constraints.
Examples include TOFU \cite{wang2020tofu}, TIFF \cite{jain2018tiff}, SemFuzz \cite{you2017semfuzz}, KCFuzz \cite{wang2021kcfuzz}, 1DVUL \cite{peng20191dvul}, SAVIOR \cite{chen2019savior};

\item More complex algorithms are adopted to enhance directedness, such as ant colony optimization, optimized simulated annealing, and particle swarm algorithm. Examples include AFLChurn \cite{zhu2021regression}, LOLLY \cite{liang2019sequence}, GREYHOUND \cite{garbelini2020greyhound};

\item To improve DGF efficiency, target unreachable inputs are filtered out in advance to save execution. Examples include FuzzGuard \cite{zongfuzzguard}, BEACON \cite{huangbeacon};

\item DGF has been used to detect specific bug types, such as memory consumption bug and algorithm complexity bug. Examples include MemLock \cite{wen2020memlock}, SlowFuzz \cite{petsios2017slowfuzz}, PERFFUZZ \cite{lemieux2018perffuzz}, MDPERFFUZZ \cite{liunderstanding}.

\end{itemize}

However, current DGF research also suffers some limitations, such as overhead deduction, equal-weighted metric bias, inflexible coordination of exploration and exploitation, source code dependence, lack of multi-object optimization, lack of multi-target coordination.
In the following sections, we will discuss the above advantages and disadvantages in detail.

\subsection{Target Identification}
According to the definition of DGF (described in Section \ref{criteria}), the fitness goal of DGF can be divided into two categories: directed for target locations, and directed for targeted bugs. Among the tool we investigated, 69\% are directed by target locations and 31\% are directed by target bugs. 

\subsubsection{Target Locations}
%Among the works we investigated, about 43\% (12/28) try to optimize how the targets are identified.

A barrier to most directed fuzzing strategies is the need for PUT target pre-labelling\cite{bohme2017directed, chen2018hawkeye, wustholz2019targeted, ye2020rdfuzz, wang2020tofu}. Manual labeling relies on the prior knowledge of the target sites, such as the line number in the source code or the virtual memory address at the binary level, to label the target and steer the execution to the desired locations. According to our statistics, 11 out of the 29 tools that target for locations need manual target labeling. However, obtaining such prior knowledge is challenging, especially for the binary code. In order to set target sites reasonably and effectively, researchers use auxiliary metadata, such as code changes from git commit logs \cite{you2017semfuzz}, information extracted from bug traces \cite{nguyen2020binary}, semantics from CVE vulnerability descriptions \cite{wangnot,  kim2019poster, garbelini2020greyhound}, or deep learning models \cite{li2019v, zhao2019suzzer, zhu2020defuzz}. Such auxiliary metadata can help identify vulnerable functions \cite{li2019v, zhao2019suzzer, kim2019poster, wangnot, zhu2020defuzz}, critical sites \cite{wang2021kcfuzz}, syntax tokens \cite{li2020gtfuzz}, sanity checks \cite{osterlundparmesan, chen2019savior}, and patch-related branches \cite{peng20191dvul,zhu2021regression} in the code and set such vulnerable code parts or sites as targets. Nevertheless, such target identification schemes still rely on additional efforts to process the information and mark the target on the PUT. It is unsuitable when fuzzing a PUT for the first time or when well-structured information is unavailable. 

To improve automation, static analysis tools \cite{christakis2016guiding, du2019leopard, situ2019energy, wustholz2019targeted, osterlundparmesan, liang2020sequence} are used to automatically find potential dangerous areas in the PUT. However, these tools are often specific to the bug types and programming languages used~\cite{osterlundparmesan}. Another direction leverages compiler sanitiser passes (e.g., UBSan \cite{ubsan}) to annotate potential bugs in the PUT~\cite{osterlundparmesan, chen2019savior}, or uses binary-level comparison (e.g., Bindiff \cite{bindiff}) to identify patch-related target branches \cite{peng20191dvul}. Deep-learning methods have been used to predict potentially vulnerable code at both binary \cite{li2019v, zhao2019suzzer} and abstract syntax tree level \cite{zhu2020defuzz}. Finally, attack surface identification components \cite{du2018towards} have also been used to identify vulnerable targets for DGF automatically.

\subsubsection{Target Bugs}

Most of the DGF tools are designed for functional goals, such as AFLGo, they need to label the potential buggy locations as target sites, and the fitness metrics are designed for approaching the target sites. Such a scheme is suitable for detecting memory corruption bugs with obvious crashes, such as overflow bugs. Among the tools we investigated, 81\% are designed for functional goals.
However, DGF tools can also detect non-functional goals. For this purpose, they need not label target sites, and the fitness metrics are designed to trigger such non-functional behavior. Among the tools we investigated, 19\% are designed for non-functional goals.
For example, UAFuzz \cite{nguyen2020binary} and UAFL \cite{wang2020uafl} leverage target operation sequences instead of target sites to find use-after-free vulnerabilities whose memory operations (e.g., allocate, use, and free memory) must be executed in a specific order. 
Memlock \cite{wen2020memlock} uses memory usage as the fitness goal to find uncontrolled memory consumption bugs. IJON \cite{aschermannijon} leverages annotations from a human analyst to overcome significant roadblocks in fuzzing and find deep-state bugs. 
%AFL-HR~\cite{medicherla2020fitness} triggers difficult-to-manifest buffer overflow and integer overflow bugs via a co-evolutionary approach. 
RVFUZZER \cite{kim2019rvfuzzer} targets input validation bugs in robotic vehicles. 
GREYHOUND \cite{garbelini2020greyhound} directs a Wi-Fi client to exhibit anomalous behaviors that deviate from Wi-Fi protocols. 
PERFFUZZ \cite{lemieux2018perffuzz} generates pathological inputs to trigger algorithmic complexity vulnerabilities \cite{petsios2017slowfuzz, lemieux2018perffuzz}. 
For the type of DGF tools that target specific bug types, since they do not need to label the target in the PUT, fuzzer can identify and trigger such bugs automatically in an evolutionary way.

\subsubsection{Distribution of Different Targets}

\begin{figure}
\centering
\includegraphics[width=0.7\columnwidth]{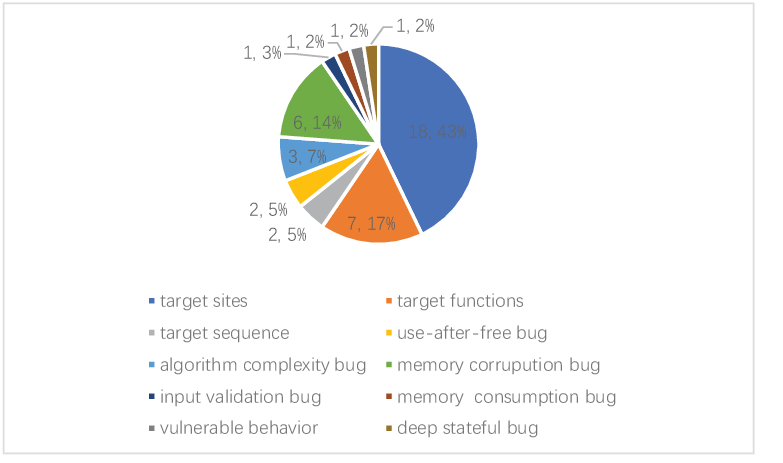}
\caption{Distribution of different targets among the tools we investigated.}
\label{target_distribution}
\end{figure}

Based on the introduction of target identification in the previous subsections, we use Figure \ref{target_distribution} to show the distribution of different targets among the tools we investigated. 
For each category in Figure \ref{target_distribution}, the number before the comma indicates the absolute number, while the number after the comma represents the percentage. We can see that tools directed for target sites account for the most (43\%), 
followed by tools directed for target functions, accounting for 17\%.  For tools directed by bug types, various bug types were considered, among them, memory corruption bug still accounts for the most (14\%). Thus, we can conclude that target locations (sites or functions) directed greybox fuzzing is still the mainstream research of DGF.

%To sum up, DGF has evolved from reaching target locations to hunting complex deep behavioral bugs,
\subsection{Fitness Metrics}
\label{metric}
The crux of DGF is using a fitness metric to measure how the current fuzzing status matches the fitness goal, so as to guide the evolutionary process. We summarize the following fitness metrics used in DGF. 

\subsubsection{Distance}
\label{bias}
\begin{figure}
\centering
\includegraphics[width=0.35\columnwidth]{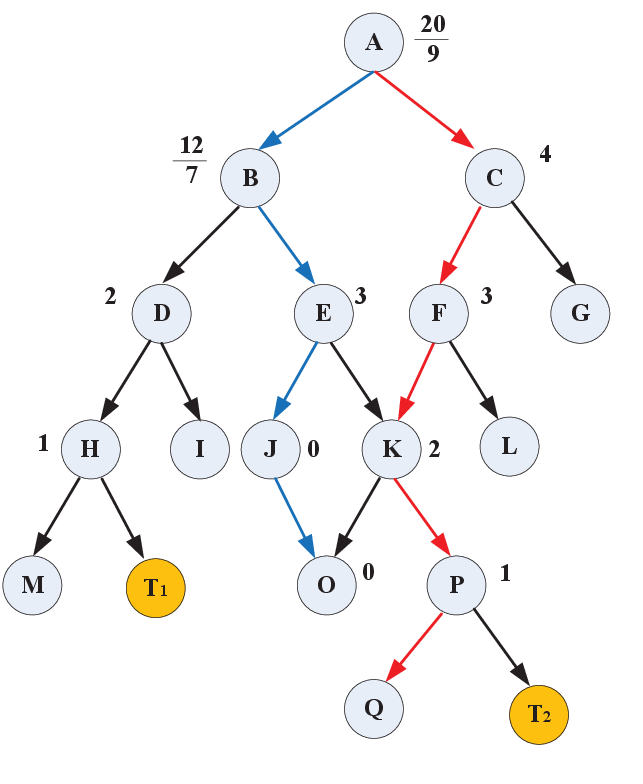}
\caption{Illustration of the distance metric.}
\label{distance}
\end{figure}

Based on our investigation, 31\% (13/42) of the directed greybox fuzzers follow the scheme of AFLGo by using the distance between the input and the target as the fitness metric. AFLGo \cite{bohme2017directed} instruments the source code at compile-time and calculates the distances to the target basic blocks by the number of edges in the call and control-flow graphs of the PUT. Then at run-time, it aggregates the distance values of each basic block exercised to compute an average value to evaluate the seed. It prioritizes seeds based on distance and gives preference to seeds that are closer to the target. 
We use the example in Figure \ref{distance} to illustrate how the distance-based metric works. In Figure \ref{distance}, each node represents a basic block, T{\footnotesize 1} and T{\footnotesize 2} are the target basic blocks. The number beside each node indicates the harmonic mean of the distances from each basic block to the target basic blocks. We can calculate the global distances of two execution paths. The distance of path A$\rightarrow$B$\rightarrow$E$\rightarrow$J$\rightarrow$O is (20/9 + 12/7 + 3)/3 $\approx$ 2.31, while the distance of path A$\rightarrow$C$\rightarrow$F$\rightarrow$K$\rightarrow$P$\rightarrow$Q is  (20/9 + 4 + 3 + 2 + 1)/5 $\approx$ 2.44. Since  $d_ABEJO$ < $d_ACFKPQ$, the seed that corresponds to path A$\rightarrow$B$\rightarrow$E$\rightarrow$J$\rightarrow$O will be prioritized.

Some follow-ups also update this scheme.
TOFU's distance metric is defined as the number of correct branching decisions needed to reach the target \cite{wang2020tofu}. 
RDFuzz~\cite{ye2020rdfuzz} combines distance with execution frequency of basic blocks to prioritize seeds. 
UAFuzz \cite{nguyen2020binary} uses a distance metric of call chains leading to target functions that are more likely to include both allocation and free functions to detect complex behavioral use-after-free vulnerabilities.
Different from using equal-weighted basic blocks in the traditional distance calculation, 
AFLChurn \cite{zhu2021regression} assigns numerical weight to a basic block based on how recently or how often it has been changed, WindRanger \cite{Du2022WindRanger} takes into account deviation  basic blocks (i.e., basic blocks where the execution trace starts to deviate from the target sites) when calculating distance.
One drawback of the distance-based method is that it only focuses on the shortest distance, and thus longer options might be ignored when there is more than one path reaching the same target, leading to a discrepancy. An example of this problem is depicted in Section \ref{bias}.
Another shortcoming is the considerable time cost when calculating the distance at the basic block level. On some target programs, users have reported that it can take many hours just to compute the distance file. For example, AFLGo spent nearly 2 hours compiling and instrumenting cxxfilt (Binutils) to generate the distance file, which is a non-negligible time cost.

\subsubsection{Similarity}

The similarity is a metric that was first proposed by Chen \textit{et al.} in Hawkeye \cite{chen2018hawkeye}, which measures the similarity between the execution trace of the seed and the target execution trace on the function level. The intuition is that seeds covering more functions in the “expected traces” will have more chances to mutate and reach the targets.

\label{bias}
\begin{figure}
\centering
\includegraphics[width=0.4\columnwidth]{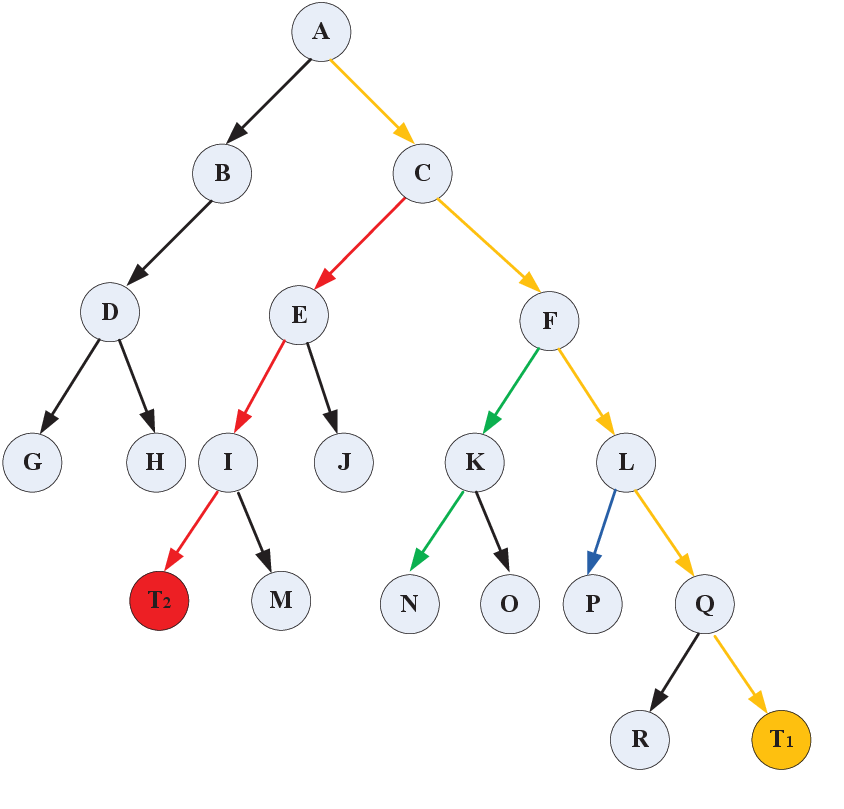}
\caption{Illustration of the similarity metric.}
\label{similarity}
\end{figure}

We use the execution tree example in Figure \ref{similarity} to illustrate how the metric of similarity works. Suppose node T{\footnotesize 1} is the target basic block, and execution trace A$\rightarrow$C$\rightarrow$F$\rightarrow$L$\rightarrow$Q$\rightarrow$T{\footnotesize 1}  is the expected trace to target T{\footnotesize 1}. There are two execution traces, A$\rightarrow$C$\rightarrow$F$\rightarrow$K$\rightarrow$N and A$\rightarrow$C$\rightarrow$F$\rightarrow$L$\rightarrow$P. We can say that trace A$\rightarrow$C$\rightarrow$F$\rightarrow$L$\rightarrow$P  is more similar to the expected trace A$\rightarrow$C$\rightarrow$F$\rightarrow$L$\rightarrow$Q$\rightarrow$T{\footnotesize 1}  than trace A$\rightarrow$C$\rightarrow$F$\rightarrow$K$\rightarrow$N. This is because trace A$\rightarrow$C$\rightarrow$F$\rightarrow$L$\rightarrow$P covers four basic blocks that are overlapped with the expected trace, which is more than trace A$\rightarrow$C$\rightarrow$F$\rightarrow$K$\rightarrow$N (only covers three). Thus, trace A$\rightarrow$C$\rightarrow$F$\rightarrow$L$\rightarrow$P is regarded as closer to the target than trace A$\rightarrow$C$\rightarrow$F$\rightarrow$K$\rightarrow$N.

Hawkeye \cite{chen2018hawkeye} combines the basic block trace distance with covered function similarity for the seed prioritization and power scheduling. LOLLY \cite{liang2019sequence} uses a user-specified program statement sequence as the target and takes the seed's ability to cover target sequences (i.e., sequence coverage) as a metric to evaluate the seed. Berry \cite{liang2020sequence} upgraded LOLLY by taking into account the execution context of target sequences. This enhances the target sequences with “necessary nodes” and uses the similarity between the target execution trace and the enhanced target sequence to prioritize the seeds. The similarity is then enriched to cover other target forms, such as operations, bug traces, and labeled locations. Formally, the similarity is \textit{the degree of overlap between the current status and target status of a certain metric, where the metric includes the length of bug traces, and the number of covered locations, covered operations, or covered functions}.

UAFL \cite{wang2020uafl} uses operation sequence coverage to guide the test case generation to progressively cover the operation sequences that are likely to trigger use-after-free vulnerabilities.
UAFuzz \cite{nguyen2020binary} also uses a sequence-aware target similarity metric to measure the similarity between the execution of a seed and the target use-after-free bug trace. 
SAVIOR \cite{chen2019savior} prioritizes seeds that have higher potentials to trigger vulnerabilities based on the coverage of labels predicted by UBSan \cite{ubsan}.
TortoiseFuzz \cite{wangnot} differentiates edges that are closely related to sensitive memory operations and prioritizes seeds based on the sensitive edge hit count in their execution paths.

For comparison, similarity-based metrics are better able to handle multi-target fitting than distance-based alternatives. 
This is because the distance-based metric is designed primarily for single-target. When there are multiple targets, a distance-based metric would calculate the distances to different targets one by one, which is less efficient. However, the similarity-based metric can handle multiple targets at one time. For example, in Figure \ref{similarity}, when analyzing execution trace A$\rightarrow$C$\rightarrow$F$\rightarrow$K$\rightarrow$N, we can calculate its similarity (i.e., the number of overlapped basic blocks) with the expected traces to target T{\footnotesize 1} and T{\footnotesize 2} simultaneously, which is more efficient.
Furthermore, similarity-based metrics can include the relationships between targets, such as the ordering of the targets \cite{nguyen2020binary}. Finally, a distance-based metric is measured at the basic block level, which would introduce considerable overheads, while a similarity-based metric can be extracted from a relatively high level to improve overall efficiency.

\subsubsection{Vulnerability Prediction Models}

\label{bias}
\begin{figure}
\centering
\includegraphics[width=0.5\columnwidth]{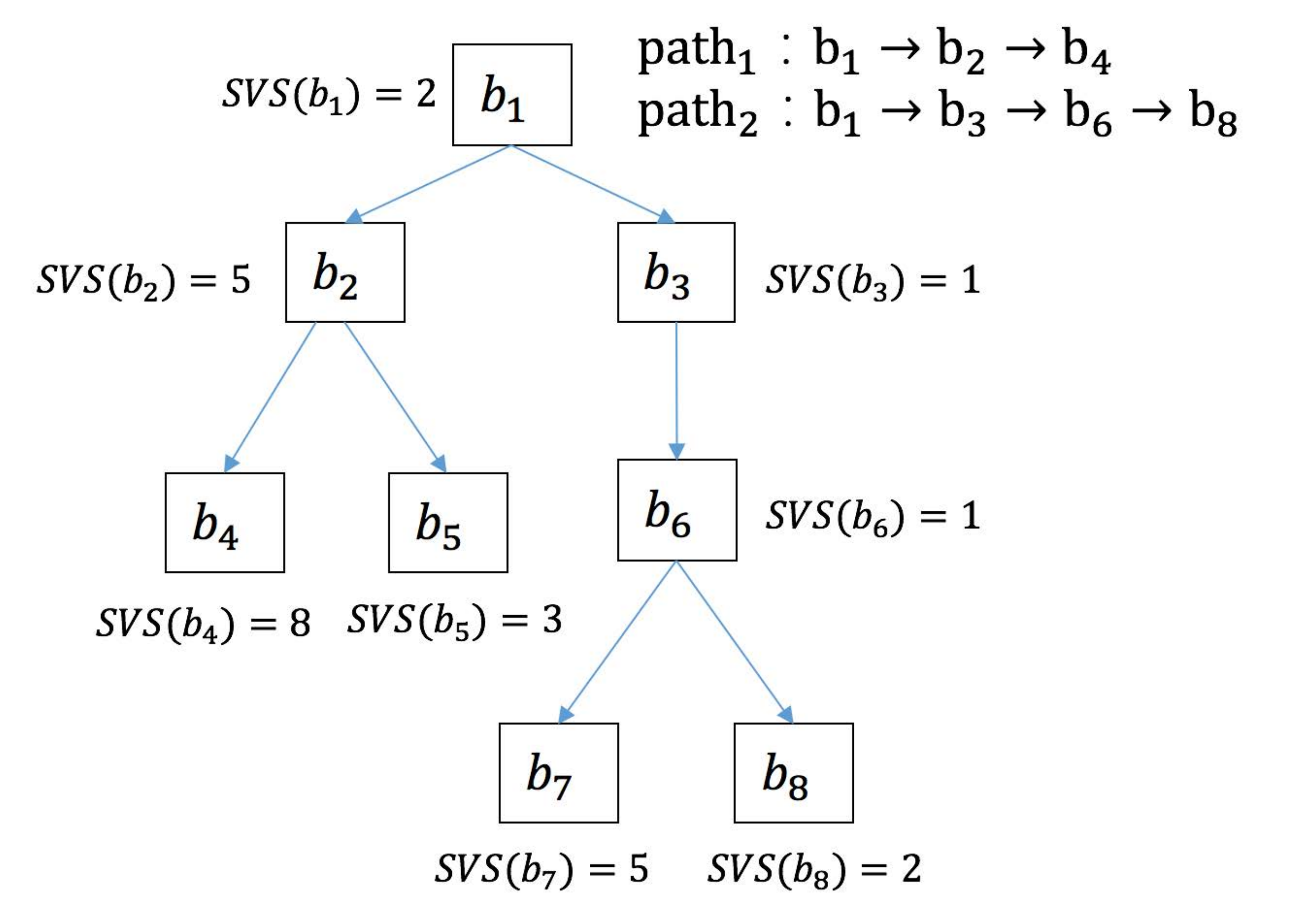}
\caption{Illustration of the vulnerable score metric.}
\label{vfuzz}
\end{figure}

Researchers also use vulnerability prediction models to quantify how likely a seed can reach a target.
%Probability is another useful metric that quantifies how likely it is that a seed can reach a target using a prediction model based on deep learning. 
Using a deep learning-based model, the vulnerable probability of a function can be predicted and each basic block in the vulnerable function is given a \textit{Static Vulnerable Score} to measure the vulnerable probability. Then for each input, the sum of the static vulnerable score of all the basic blocks on the execution path is used as a fitness score to prioritize inputs with higher scores \cite{li2019v,zhao2019suzzer}. 
Figure \ref{vfuzz} illustrates how the vulnerable score metric works in V-Fuzz \cite{li2019v}. SVS is the static vulnerable score for each basic block predicted by the deep learning-based model.
We assume there are two inputs $i_1$, $i_2$, and the execution paths of the two inputs are $path_1$ and $path_2$ respectively. 
Suppose $path_1$ is $b_1 \rightarrow b_2 \rightarrow b_4$, and $path_2$ is $b_1 \rightarrow b_3 \rightarrow b_6 \rightarrow b_8$. 
The fitness score of input $i_1$, and $i_2$ are $f_1$ and $f_2$ respectively. Then, $f_1$ = 2 + 5 + 8 = 15,
 $f_2$ = 2 + 1 + 1 + 2 = 6.  As  $f_1$ is larger than $f_2$, the input $i_1$ will be selected as a favored seed.

%V-Fuzz \cite{li2019v} and SUZZER \cite{zhao2019suzzer} predict the vulnerable probability of functions based on a deep learning-based model and give each basic block in the vulnerable function a static score. Then for each input, they calculate the sum of the static score of all the basic blocks on the execution path and prioritize the inputs with higher scores. 
TAFL \cite{situ2019energy} extracts semantic metrics of the PUT and uses static semantic analysis to label regions, including sensitive, complex, deep, and rare-to-reach regions, that have a higher probability of containing vulnerabilities and strengthens fuzzing towards such regions.
Joffe \cite{joffe2019directing} uses crash likelihood generated by a neural network to direct fuzzing towards executions that are crash-prone. 
The probability-based metric can combine seed prioritization with target identification to direct fuzzing towards potentially vulnerable locations without relying on the source code. Using deep learning models, a probability-based metric can be extended to targeting properties other than crashes, such as information leaks, exploits, as well as specific crash types, and different resource usages. Besides, deep learning methods have been proven to be able to detect several types of vulnerabilities simultaneously \cite{li2019v}. However, a major weakness is that the accuracy at present is to some extent limited.

\subsubsection{Customized Fitness Metrics}

Apart from the above categories, researchers also propose customized metrics for DGF.
W\"ustholz \textit{et al}~\cite{wustholz2019targeted} used online static look ahead analysis to determine a path prefix for which all suffix paths cannot reach a target location. Directed fuzzing is then enabled by strategically scheduling the energy of fuzzing to stress the path prefix that might reach the target locations.
KCFuzz \cite{wang2021kcfuzz} defines the parent nodes in the path to the target as keypoints and directs fuzzing using keypoint coverage.
CAFL \cite{lee2021constraint} aims to satisfy a sequence of constraints (i.e., the combination of a target site and the data conditions) instead of reaching a set of target sites. It defines the distance of constraints as how well a given seed satisfies the constraints, and  prioritizes the seeds that better satisfying the constraints in order.
AFL-HR~\cite{medicherla2020fitness} and HDR-Fuzz \cite{medicherla2021hdr} adopt a vulnerability-oriented fitness metric called \textit{headroom}, which indicates how closely a test input can expose a hard-to-manifest vulnerability (e.g., buffer or integer overflow) at a given vulnerability location. 
PERFFUZZ \cite{lemieux2018perffuzz} uses the new maxima of execution counts for all program locations as feedback to generate pathological inputs.
To systematically measure fitness, a customized fitness metric also takes into account multiple dimensions simultaneously, including basic code coverage, block weight, number of state transitions, execution time, anomaly count, and so forth \cite{ji2020aflpro, garbelini2020greyhound}. In addition, non-functional properties such as memory usage ~\cite{wen2020memlock} and control instability of robotic vehicles~\cite{kim2019rvfuzzer} can also be used to direct fuzzing.

\subsubsection{Distribution of Fitness Metrics}

\begin{figure}
\centering
\includegraphics[width=0.7\columnwidth]{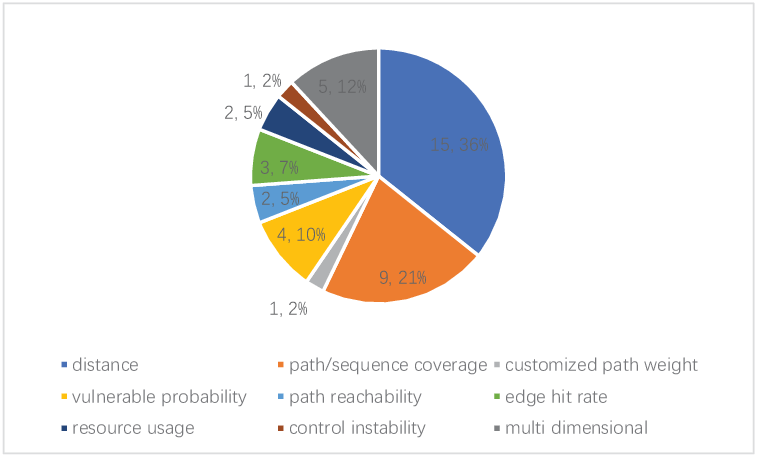}
\caption{Distribution of different fitness metrics among the tools we investigated.}
\label{fitness_distribution}
\end{figure}

Based on the introduction of fitness metrics in the previous subsections, we use Figure \ref{fitness_distribution} to show the distribution of fitness metrics among the tools we investigated. 
From Figure \ref{fitness_distribution}, we can see that the distance metric accounts for the most (36\%), 
followed by the fitness metric of path/sequence coverage, accounting for 21\%. 
Notably, as new fitness metrics, vulnerability prediction models (i.e., vulnerable probability) and customized fitness metrics (i.e., edge hit rate ) are important, accounting for 10\% and 7\%, respectively.
Besides, the multi-dimensional fitness metric is also prevalent, accounting for 12\%. 
Thus, we can conclude that the distance metric is still the major fitness metric for DGF, but new fitness metrics are growing fast.

% For example, AFLPro \cite{ji2020aflpro} defines a multi-dimensional fitness metric that takes the code coverage, local basic block weight and global path weight as core dimensions, and takes the seed length and seed execution time into consideration to select seeds. GREYHOUND \cite{garbelini2020greyhound} 

%However, the widely adopted basic block level distance calculation does not take account of the edge direction and only considers the number of edges. As a result, many infeasible edges (e.g., edges reverse to the path direction, and edges do not affect the target reachability) are included in calculating the distance, leading to bias in the seed prioritization (discuss in detail in Section \ref{bias}). Differently, a path-based seed prioritization is usually calculated on a tree, in which the path starts from the root node and links the intermediate edges in a directed manner.
%Under such a situation, when analyzing the relationship between a path and the targets, we should only care about the prefix path section before reaching a target. The suffix is meaningless once a path goes through a target as it does not affect the reachability.

\subsection{Fuzzing Optimization}

\begin{figure}
\centering
\includegraphics[width=0.5\columnwidth]{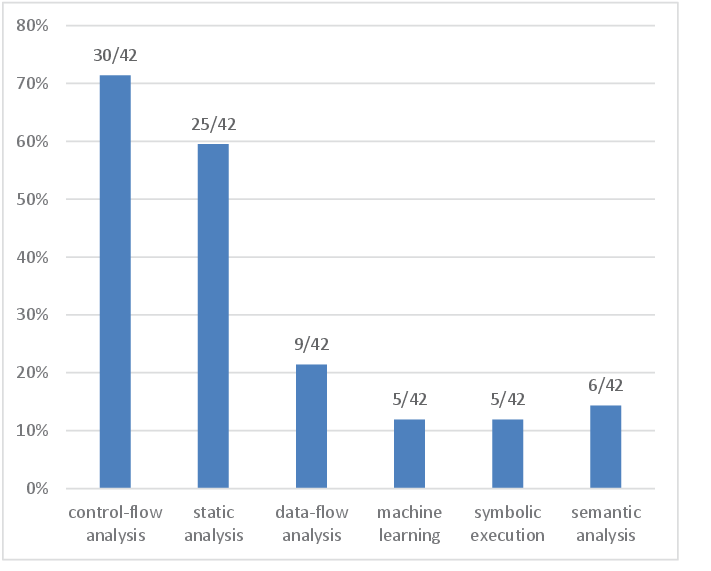}
\caption{Statistics of mainstream optimization techniques used in DGF.}
\label{statistics}
\end{figure}

Since a native fuzzer that uses randomly generated test inputs can hardly reach deep targets and is less effective at triggering deep bugs along complex paths, various program analysis techniques, such as static analysis, control-flow analysis, data-flow analysis, machine learning, semantic analysis, and symbolic execution, have been adopted to enhance the directedness of reaching corner cases and flaky bugs. Figure \ref{statistics} shows the statistics of mainstream optimization techniques used in DGF.

Among the tools investigated, 71\% of them relied on the control-flow analysis to evaluate seeds and determine the reachability to the targets; 60\% of them leverage static analysis to automatically identify targets \cite{chen2019savior} and extract information from the PUT \cite{chen2018hawkeye, wustholz2019targeted}; 
21\% use data-flow analysis (mainly taint analysis) to identify the relationship between the input and the critical program variables \cite{mathis2019parser, peng20191dvul, jain2018tiff} or to optimize mutator scheduling \cite{ wang2020uafl}; 12\% use machine learning to predict vulnerable code \cite{li2019v} and filter out unreachable inputs \cite{zongfuzzguard}; 12\% integrate symbolic (concolic) execution to solve complex path constraints \cite{kim2019poster, liang2020sequence, chen2019savior, peng20191dvul}; and finally, 14\% adopt semantic analysis to identify vulnerable targets automatically \cite{kim2019poster, situ2019energy, you2017semfuzz} and learn input fields semantics to optimize mutation. The next section will discuss the key steps of greybox fuzzing and how they are optimized for directedness.

%(7/28) support binary code.
%24/28 of the tools support multi-targets, however, only 3 of the noticed the relationship among targets. 
%14/28 tried to identify the targets automatically, 
%5/28 leverages data-flow analysis to optimize 
%9/28 input generation.
%7/28 mutation
%1/28 e-e

%static analysis 13/32
%control-flow analysis, 28/32 
%data-flow analysis, 6/32
%machine learning, 5/32
%semantic analysis, 4/32
%symbolic execution 5/32

\subsubsection{Input Optimization}
%Once the targets are marked, DGF needs to generate a seed input to invoke the fuzzing process. 

A good seed input can drive the fuzzing process closer to the target location and improve the performance of the later mutation process. According to Zong \textit{et al.}, on average, over 91.7\% of the inputs of AFLGo cannot reach buggy code~\cite{zongfuzzguard}. There are thus many opportunities to increase the ability of DGF by enhancing the input generation. Dynamic taint analysis \cite{jain2018tiff} and semantic information~\cite{you2017semfuzz} can assist in generating valid input that matches the input format ~\cite{wang2020tofu, mathis2019parser}. These techniques also increase the probability of hitting vulnerable functions~\cite{you2017semfuzz} or security-sensitive program sites, such as maximizing the likelihood of triggering memory corruption bugs~\cite{jain2018tiff}. Except that, FuzzGuard~\cite{zongfuzzguard} utilizes a deep-learning-based approach to predict and filter out unreachable inputs before exercising them, which saves time that can then be spent on real execution. BEACON \cite{huangbeacon} prunes infeasible paths (i.e., paths that cannot reach the target code at runtime) with a lightweight static analysis, which can reject over 80\% of the paths executed during fuzzing.

\subsubsection{Seed Prioritization}
The core of DGF is the prioritization of seeds (for mutation) that are closest to the targets.
DGF implementation is effectively the act of closest seed-target relation prioritization. No matter what kind of fitness metric it adopts, seed prioritization is mainly realized based on control-flow analysis. Distance-based approaches~\cite{bohme2017directed, chen2018hawkeye, wang2020tofu, ye2020rdfuzz, peng20191dvul, nguyen2020binary, osterlundparmesan} calculate the distance to the target basic blocks from the number of edges in the call and control-flow graphs of the PUT. Similarity-based approaches ~\cite{liang2019sequence, liang2020sequence, wang2020uafl, nguyen2020binary, wangnot} take the seed's ability to cover the target edges on the control-flow graph as a metric to evaluate the seed. Prediction model-based approaches \cite{li2019v, zhao2019suzzer} also rely on the attributed control-flow graph (i.e., using a numerical vector to describe the basic block in a control-flow graph, where each dimension of the vector denotes the value of a specific attribute of the basic block) to represent a binary program and extract features for deep learning. 
A further point to note is that directed hybrid fuzzing \cite{kim2019poster, liang2020sequence, chen2019savior, peng20191dvul,wang2021kcfuzz} combines the precision of DSE and the scalability of DGF to mitigate their individual weaknesses. DGF can prioritize and schedule input mutation to get closer to the targets rapidly, while DSE can reach more in-depth code by solving complex path constraints.

\subsubsection{Power Scheduling}

\begin{figure*}
\centering
\subfigure[The variations of APS Factor for seeds with a normalized distance of 0.1.]{
\includegraphics[width=0.45\columnwidth]{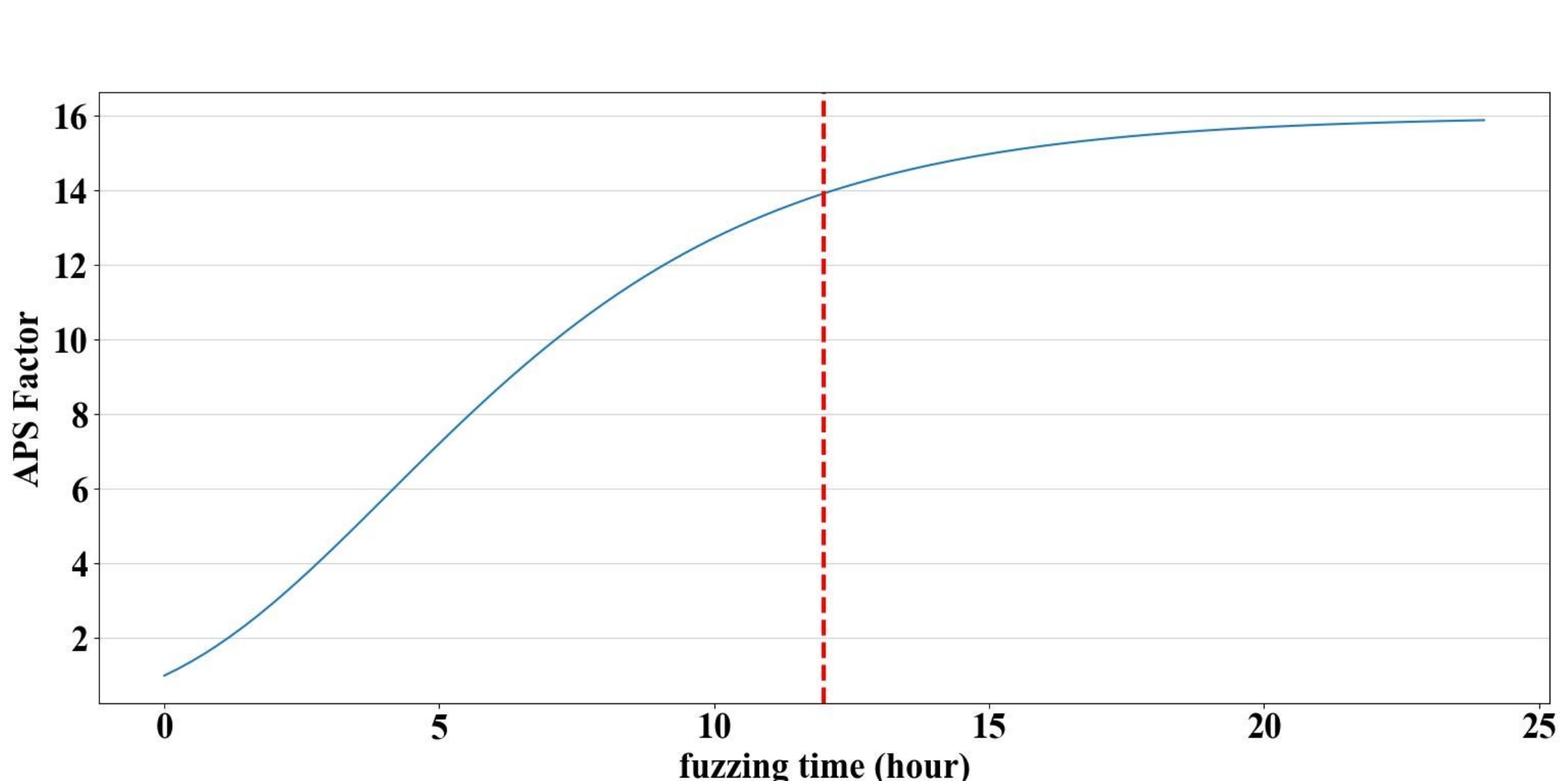}} \&
\subfigure[The variations of APS Factor for seeds with a normalized distance of 0.9.]{
\includegraphics[width=0.41\columnwidth]{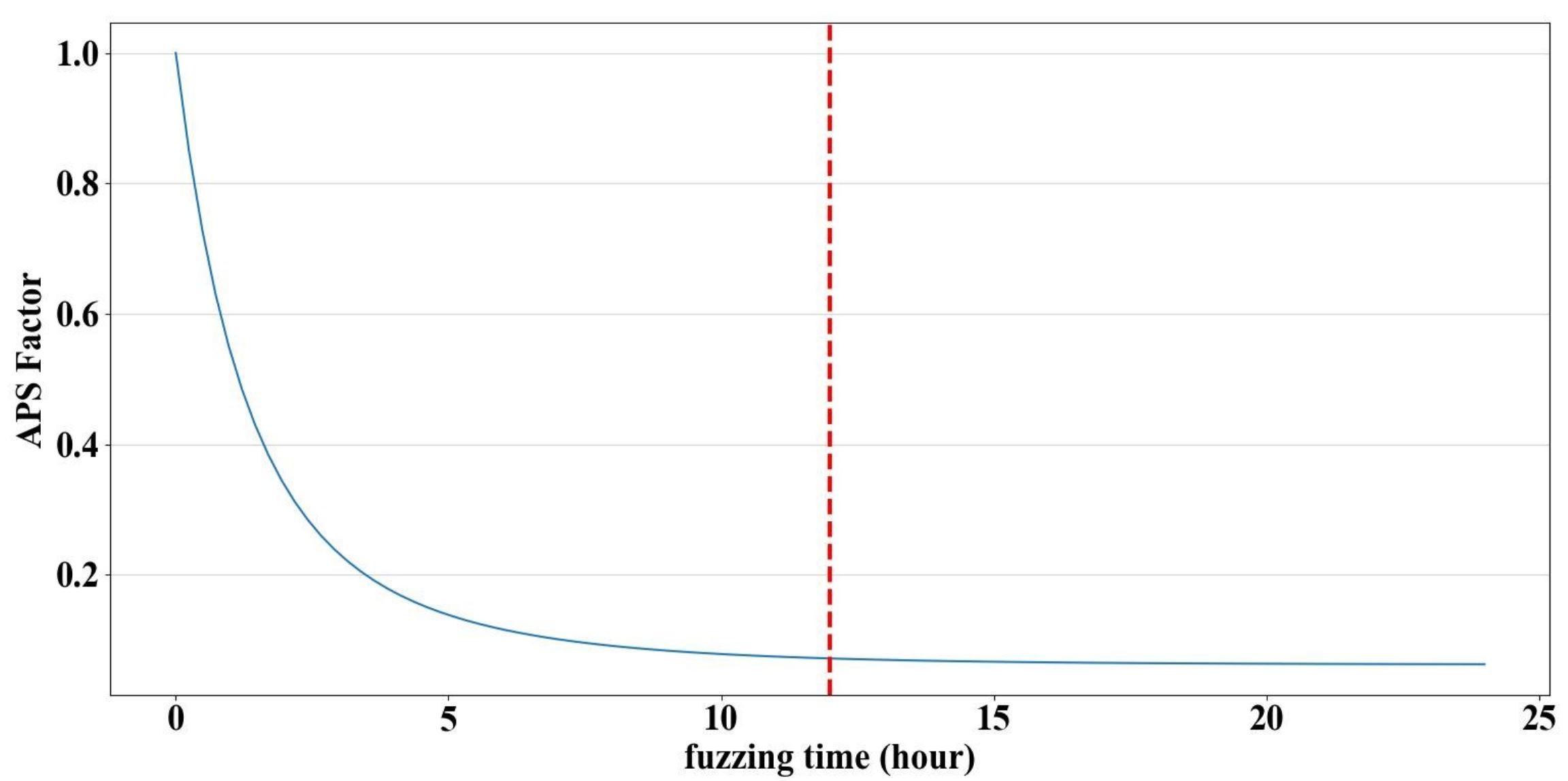}}
\caption{Illustration of the simulated-annealing-based power scheduling.}
\label{APS}
\end{figure*}

After being selected, the seeds nearest to their targets are subjected to greater fuzzing opportunities by assigning more power, i.e., more inputs are produced by mutating them. Whereas AFL uses execution trace characteristics such as trace size, PUT execution speed, and order in the fuzzing queue for power scheduling, most directed greybox fuzzers use simulated annealing to allocate energy. Unlike traditional random walk scheduling, which always accepts better solutions and may be trapped in a local optimum, simulated annealing accepts a solution that is worse than the current one with a certain probability, so it is possible to jump out of local optima and reach the globally optimal solution \cite{liang2019sequence}.

To illustrate simulated annealing-based power scheduling in a straight way, we employed AFLGo to test libxml2 and collected the simulated annealing algorithm factor (APS Factor) of seeds with normalized distances of 0.1 and 0.9. As shown in Figure \ref{APS}, the x-axis represents the fuzzing time ranging from 0 to 24 hours, and the y-axis denotes the variations of the APS Factor. Meanwhile, we set the time threshold for simulated annealing in AFLGo to 12 hours, marked by a red dotted line. Additionally, we adopted an exponential growth mode for energy for energy over time (i.e., exp mode in AFLGo).
Figure \ref{APS} (a) illustrates the increasing trend of the APS Factor with a normalized distance of 0.1 over time. As fuzzing progresses, the APS Factor with a normalized distance of 0.1 exhibits exponential growth and eventually converges to 16. On the other hand, Figure \ref{APS} (b) demonstrates the decreasing trend of the APS Factor with a normalized distance of 0.9 over time. With continuous fuzzing, the APS Factor with a normalized distance of 0.9 shows exponential decay and eventually converges to 1/16.

AFLGo \cite{bohme2017directed} was the first to use a simulated annealing-based power schedule to gradually assign more energy to seeds that are closer to the target locations while reducing energy for distant seeds. 
Hawkeye \cite{chen2018hawkeye} added prioritization to simulated annealing to allow seeds that are closer to the target to mutate first.
AFLChurn \cite{zhu2021regression} proposes a byte-level power scheduling based on ant colony optimization which assigns more energy to bytes that generate more "interesting" inputs. 
LOLLY \cite{liang2019sequence} and Berry \cite{liang2019sequence} optimized simulated annealing-based power schedules with a temperature threshold to coordinate the cooling schedule in both the exploration and exploitation stages. In the exploration stage, the cooling schedule randomly mutates the provided seeds to generate many new inputs, while in the exploitation stage, it generates more new inputs from seeds that have higher sequence coverage, which is similar to the traditional gradient descent algorithm \cite{liang2019sequence}. In addition to simulated annealing, GREYHOUND \cite{garbelini2020greyhound} also adopts a custom generational particle swarm algorithm, which is better suited for the non-linear and stochastic behavior of the protocol model.

%UAFuzz\cite{nguyen2020binary} assigns energy to a seed proportionally to the number of targets covered in sequence and with a corrective factor based on seed distance and cut-edge coverage.

\subsubsection{Mutator Scheduling}
%Some fuzzers (8 out of 28) 
Optimizing mutator scheduling is another viable way of bettering directed fuzzing. Reasonable scheduling of mutators can enhance the directedness of inputs by improving the precision and speed of seed mutation.
%In addition to seed prioritization, optimizing mutation strategies is another direction to enhance directed fuzzing.
A viable approach is to first classify mutators into different granularities, such as coarse-grained and fine-grained \cite{chen2018hawkeye, li2019v, you2017semfuzz, situ2019energy}, and then dynamically adjust them according to the actual fuzzing states. Coarse-grained mutators are used to change bulk bytes during mutations to move the execution towards the ``vulnerable functions'', while fine-grained only involves a few byte-level modifications, insertions, or deletions,  so as to monitor the ``critical variables'' \cite{you2017semfuzz}. The fuzzer gives a lower chance of coarse-grained mutation when a seed can reach the target function. Once the seed reaches targets, the time for fine-grained mutations increases as coarse-grained mutations decrease. In practice, the scheduling of mutators is controlled by empirical values \cite{chen2018hawkeye, li2019v}. Situ \textit{et al.} ~\cite{situ2019energy} gives two empirical observations—that (1) coarse-grained mutators outperform fine-grained mutators on path growth; and (2) the use of multiple mutations offers improved performance compared to each individual mutation.

%ProFuzzer \cite{you2019profuzzer} entails different mutation policies according to the input field types recognized by input type probing.

\subsubsection{Mutation Operations}
Data-flow analysis, such as taint analysis, can reflect the effect of the mutation in the generated inputs, thus, it is helpful to optimize both mutation operations and input generation.
RDFuzz~\cite{ye2020rdfuzz} leverages a disturb-and-check method to identify and protect ``distance-sensitive content'' from the input, i.e., the critical content to maintain the distance between the input and the target, and once altered, the distance would become larger. Protecting such content during mutation can help to approach the target code location more efficiently.  
UAFL \cite{wang2020uafl} adopts information flow analysis to identify the relationship between the input and the program variables in the conditional statement. It regards input bytes that are more likely to change the values of target statement as with higher ``information flow strength'', and assigns higher mutation possibility for them. The higher the information flow strength, the stronger this byte influences the values of the variables.
SemFuzz \cite{you2017semfuzz} tracks the kernel function parameters that the critical variables depend on via backward data-flow analysis.
%SeededFuzz \cite{wang2016seededfuzz} utilizes dynamic taint analysis to identify the bytes of seeds that can influence values at security-sensitive program sites.
TIFF \cite{jain2018tiff} infers input type by type-based mutation to increase the probability of triggering memory corruption vulnerabilities. It leverages in-memory data-structure identification to identify the types of each memory address used by the application and uses dynamic taint analysis to map what input bytes end up in what memory locations.
Nevertheless, data-flow analysis usually enlarges the run-time overhead.

%However, compared to control-flow analysis, data-flow analysis is less prevalent in DGF because it usually enlarges the run-time overhead.
\subsection{Base Tools}

\begin{figure}
\centering
\includegraphics[width=0.7\columnwidth]{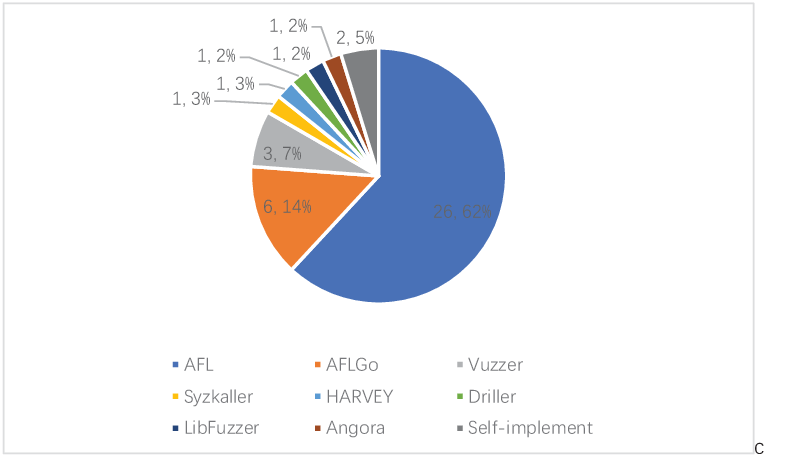}
\caption{Distribution of the base tools.}
\label{base_distribution}
\end{figure}

Based on the statistics in Table 1, we use Figure \ref{base_distribution} to show the distribution of the base tools.
As Figure \ref{base_distribution} shows, 62\% (20/42) of the DGF tools are built on AFL, which is similar to the situation of CGF. As AFL is a well-structured framework with good performance, it is suitable for further development, including DGF research. Most DGF tools with customized fitness metrics would adopt AFL as the base tool. The second most used (6/42) base tool is AFLGo. Since AFLGo is the first DGF tool based on the basic block distance, it is suitable for tools with distance-based fitness metrics. In addition, some base tools are adopted owing to the test requirement. For example, Syzkaller is used as the base tool for DGF in the kernel, VUzzer is used as the base tool for DGF at the binary level, and HARVEY is used as the base tool for testing Ethereum smart contracts.

There are also tools built without a based tool, such as TOFU and RVFuzzer. For TOFU, though it uses a distance-based fitness metric, the author did not mention any base tools in their paper, but stated that “the high-level structure of TOFU is similar to that of AFL”.  Since TOFU is not open-sourced, we can only infer that TOFU is self-devised. The reason may be that it relies on the compile tool WLLVM to extract ICFG, and WLLVM is incompatible with the commonly used base tools. As for the RVFuzzer, since it is used to test robotic vehicles, most of the existing base tools, such as AFL and AFLGo, can not be used directly due to the source code requirement and environmental requirements. In such a situation, the author would develop a customized tool, which is more usable, more lightweight, and more efficient for the target system.
  
The main advantage of using base tools is the convenience of implementing tools. The already widely used base tool can provide a reliable platform to build new tools. For example, when developing a distance-based DGF tool based on AFL, we only need to modify a few modules (e.g., the fitness metric) in the framework, which is very time-saving. However, the major disadvantage lies in the compatibility. When the testing target has restrictions, such as only having binary code or has different running environment, the commonly used base tools are not suitable or less efficient. A self-devised customized tool is needed, which is more usable, more lightweight, and more efficient for the target environment. For example, RVFuzzer for testing robotic vehicles.

\section{Challenges Faced by Directed Greybox Fuzzing}

%The following discussion provides the difficulties faced by DGF, which is built from consideration of the latest work into directed greybox fuzzers.

\subsection{Performance Deduction}
To realize directedness in fuzzing, most researchers use additional instrumentation and data analysis, for example, by static analysis, symbolic execution, taint analysis, and machine learning. However, such additional analysis inevitably incurs performance deduction. The analysis cost is one of the biggest drawbacks of almost all the directed fuzzing techniques.
To evaluate the performance of directed greybox fuzzers, researchers usually focus on the ability to reach targets or expose bugs, using metrics such as \textit{Time-to-Reach} (the length of the fuzzing campaign until the first testcase that reaches a given location) or \textit{Time-to-Exposure} (the length of the fuzzing campaign until the first testcase that exposes a given error \cite{bohme2017directed}) to measure the performance of directed greybox fuzzers, while ignoring the measurement of performance overhead. However, for a given fuzzing time budget, higher efficiency means more fuzzing executions and consequently, more chance to reach the target. Thus, optimizing efficiency is a major challenge to improve directedness. Based on our investigation, we summarize the following solution to improve DGF efficiency.
\begin{itemize}

\item [-]\textbf{Move the heavy execution-independent computation from run-time to compile-time.} For example, AFLGo measures the distance between each basic block and a target location by parsing the call graph and intra-procedure control flow graph of the PUT. Since both parsing graphs and calculating distances are very time-consuming,
 AFLGo moves most of the graph parsing and distance calculation to the instrumentation phase at compile-time in exchange for efficiency at run-time. Such compile-time overhead can be saved when a PUT is tested repeatedly;

\item[-]\textbf{Filter out the unreachable inputs to the target before execution.} For example, FuzzGuard~\cite{zongfuzzguard} utilizes a deep-learning-based approach and BEACON \cite{huangbeacon} uses a lightweight static analysis to find infeasible paths to the targets in advance, which can save over 80\% of the path execution during fuzzing;

\item[-]\textbf{Use more light-weight algorithms. }For example, AFLChurn \cite{zhu2021regression} leverages light-weight ant colony optimization instead of expensive taint analysis to find ``interesting bytes'' and realize a byte-level power scheduling;

\item[-]\textbf{Leverage parallel computing.} For example, HDR-Fuzz  \cite{medicherla2021hdr} uses another core to run AddressSanitizer in parallel and provides guidance to the directedness. Large-scale parallel fuzzing  \cite{chen2019enfuzz, liang2018pafl} can also be adopted to improve efficiency further.

\end{itemize}

However, these approaches might be less effective for certain situations. For example, for the continuously evolving software, moving analysis from run-time to compile-time may not get obvious benefit as it will have to be re-run after every single code change during compile time. Filtering out unreachable inputs and pruning infeasible paths inevitably introduce false reports, making unreachable inputs and infeasible paths remain, and such computational-complex paths and inputs would influence the performance.

In recent years, two state-of-the-art techniques, namely BEACON \cite{huangbeacon} and SelectFuzz \cite{luo2023selectfuzz}, have significantly improved the speed of bug exposure by reducing the cost of fuzzing. BEACON leverages symbolic execution to analyze the feasibility of different paths and eliminates those that cannot lead to the target, thereby reducing the overall fuzzing cost. In a similar vein, SelectFuzz conducts a preliminary analysis of the reachability of basic blocks and selectively instruments and calculates seed distances for blocks that are reachable. As a result, the overhead associated with instrumentation and seed distance calculation is minimized.  By this means, SelectFuzz \cite{luo2023selectfuzz} avoids exploring irrelevant code, further reducing the cost of fuzzing. By adopting these strategies, both BEACON \cite{huangbeacon} and SelectFuzz \cite{luo2023selectfuzz} effectively decrease the fuzzing cost by minimizing the exploration of code that does not contribute to reaching the desired targets, thereby enhancing the speed of bug exposure.
In contrast, other techniques such as Windranger \cite{Du2022WindRanger}, CAFL \cite{lee2021constraint}, and FuzzGuard \cite{zongfuzzguard} actually increase the cost of fuzzing due to their requirements for analyzing and collecting DBBs and path constraints, or collecting and filtering seeds. However, in our evaluation, we found that the additional fuzzing cost incurred by these techniques does not have a significant impact on the overall fuzzing throughput. Furthermore, the fitness metrics and fuzzing strategies proposed by these techniques can effectively guide DGF to reach targets faster. Considering the improved speed of bug exposure achieved by these fitness metrics, the additional cost of fuzzing is deemed acceptable. This has achieved a tradeoff between fuzzing cost and bug exposure.

\subsection{Equal-weighted Metrics Bias Seed Prioritization}
\label{probability}

In most of the state-of-the-art directed greybox fuzzers, the seed prioritization is based on equal-weighted metrics, i.e., treat each branch jump in the control-flow graph as having equal probability. Taking the widely used distance-based metric as an example, where the distance is represented by a number of edges, namely the transitions among basic blocks. However, such measurement ignores the fact that different branch jumps have different probabilities to take, and thus, biases the performance of directed fuzzing.

\begin{figure}
\centering
\includegraphics[width=0.3\columnwidth]{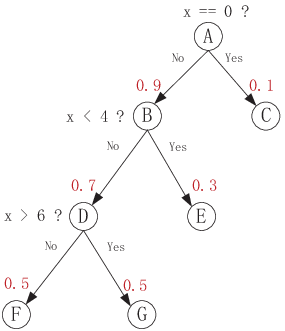}
\caption{Equal-weighted metric incurs bias in distance-based seed prioritization.}
\label{probability_bias}
\end{figure}

Fig.~\ref{probability_bias} shows a control-flow graph fragment of a simple example to illustrate the problem. Suppose input \textit{x} is an integer ranging from 0 to 9. Obviously, the probability of jumping from node A to node C is 0.1, and from node A to node B is 0.9. We can also compute the probabilities of other jumps by the branch conditions. %It is also possible to determine all other likelihoods herein by assessing the state of each branch. 
When using a distance-based metric, the distance of A $\rightarrow$ C is shorter than that of A $\rightarrow$ G because A $\rightarrow$ C has only one jump but A $\rightarrow$ G has three jumps. However, when taking the jump probability into account, the probability of A $\rightarrow$ C is 0.1, while the probability of A $\rightarrow$ G is 0.9 $\times$ 0.7 $\times$ 0.5 $\approx$ 0.3, which is more likely to be taken than A $\rightarrow$ C and should be considered as ``shorter''. Thus, it is reasonable to also consider the weight difference when designing the fitness metric.
Though, this is a hypothetical example, such a problem is realistic and frequent in the real-world program. 
One common case is when A $\rightarrow$ C represents an execution path through the error-handling code. The error-handling code is usually short and simple, which is used to retrieve resources, such as free the allocated memory. Thus, the execution path through the error-handling code to the target is usually short in distance (e.g., one jump). However, since error-handling code is rarely executed, such an execution path has a low probability. If we only consider distance, the path through the error-handling code would be over-emphasized, and we would ignore the bug-prone regular code, leading to a bias.
%Thus, it is reasonable to also consider the weight difference when designing the fitness metric.

%Here, it is argued that one must also take the branch jump probability into account when constructing the fitness metric.
One solution is taking branch jump probability into account to construct weighted fitness metrics. In that case,
each seed is prioritized by the probability of converting the current execution path to a target path that goes through the target. Since an execution path can be viewed as a Markov Chain of successive branches~\cite{bohme2017coverage}, the path probability can be calculated by gathering the probabilities of all the branches within the path. Then the branch probability can be statistically estimated by calculating the ratio based on the Monte Carlo method \cite{wangnot}. By its very nature, the randomness and high throughput of the fuzzing process fulfill the requirements for random and large sampling with the Monte Carlo method. Thus, the distribution density can formally estimate the branch jump probability in a lightweight fashion.

%One solution to avoid bias caused by equal-weight metrics is using probability-based metrics.
%The density of the stationary distribution formally describes the likelihood that the fuzzer exercises a specific path after a certain number of iterations.
%A Monte Carlo based method requires two conditions: 1) the sampling should be random; 2) the sample scale should be large~\cite{zhao2019send}. Fortunately, 

One possible drawback of evaluating the reachability of the target based on probability is the potential run-time overhead. Both the statistical jump counting and the probability calculation introduce extra computation. One way to alleviate performance deduction is interval sampling. Compressing the volume of jump statistics by appropriate sampling can accelerate the probability calculation and alleviate the space requirement for storage. Another way is to accelerate how the meta-data of jump statistics is stored and accessed. 
On one hand, the probability-based approach updates the jump statistics very often and the reachability judgement also requires a quick edge tracing. On the other hand, since a jump usually has only two branches, the data distribution (e.g. based on a matrix) would be relatively sparse, which dramatically increases space consumption. Thus, a customized data structure that balances the time and space complexities is required. 

%Conventionally, graph-based data is stored in an adjacency table. However, since the probability-based approach updates the jump statistics very often and the reachability judgment also requires a quick edge tracing, the adjacency table is unsuitable due to its low efficiency when accessing data. Another option is the adjacency matrix \cite{wangnot}, which supports quick data access. However, since a jump usually has two branches, the matrix would be vast, yet the data distribution remains relatively sparse, which dramatically increases space consumption. Therefore, a pre-condition to leverage a probability-based approach is designing a customized data structure that balances the time and space complexities.

%Du \textit{et al.} \cite{du2020detection} improve DGF efficiency by identifying and providing more suspiciously malicious basic blocks that are feature similar to the target blocks.

\subsection{Global Optimum Discrepancy in the Distance-based Metric}
\label{bias}
\begin{figure}
\centering
\includegraphics[width=0.5\columnwidth]{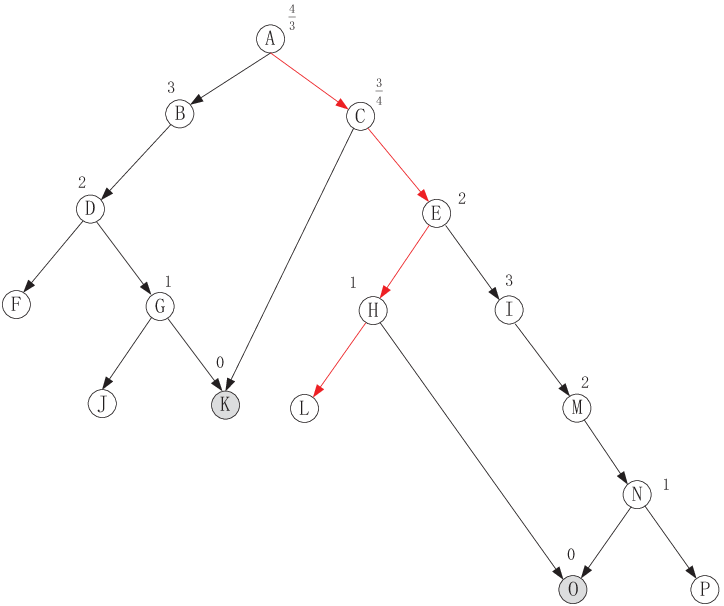}
\caption{Discrepancy introduced by distance-based seed prioritization metric.}
\label{globalbias}
\end{figure}

%The coordinated targeting of multiple aims within directed fuzzing is a critical objective. 
When measuring multiple targets with a distance-based metric, one strategy is to seek the global shortest distance between the execution path and all the targets using Dijkstra’s algorithm~\cite{bohme2017directed, chen2018hawkeye, wustholz2019targeted, ye2020rdfuzz, wang2020tofu}. However, such global optimum might miss local optimal seeds that are closest to a specific target, leading to a discrepancy. In order to elucidate this case, an example is depicted in Fig.~\ref{globalbias}.
In this control-flow graph fragment, node K and O are the target nodes. For the three seeds under test, one exercises path A$\rightarrow$B$\rightarrow$D$\rightarrow$G$\rightarrow$K, one exercises path A$\rightarrow$C$\rightarrow$E$\rightarrow$I$\rightarrow$M$\rightarrow$N$\rightarrow$O, and the last exercises path A$\rightarrow$C$\rightarrow$E$\rightarrow$H$\rightarrow$L. Based on the distance formula defined by B\"ohme \textit{et al}.~\cite{bohme2017directed}, the harmonic distances were calculated between each node in the three paths to the two targets—these are labelled in the figure. The global distance for each of the three seeds are $d_{ABDGK} = (4/3 + 3 + 2 + 1 + 0)/5 \approx 1.47$, $d_{ACEIMNO} = (4/3 + 3/4 + 2 + 3 + 2 + 1 + 0)/7 \approx 1.44$, and $d_{ACEHL} = (4/3 + 3/4 + 2 + 1)/4 \approx 1.27$. Since $d_{ACEHL}$ is the smallest of the three, one should prioritize the seed for path A$\rightarrow$C$\rightarrow$E$\rightarrow$H$\rightarrow$L. However, this is unreasonable because path A$\rightarrow$B$\rightarrow$D$\rightarrow$G$\rightarrow$K goes through target node K, while path A$\rightarrow$C$\rightarrow$E$\rightarrow$I$\rightarrow$M$\rightarrow$N$\rightarrow$O goes through target O, but path A$\rightarrow$C$\rightarrow$E$\rightarrow$H$\rightarrow$L does not reach any targets. Intuitively, as path A$\rightarrow$C$\rightarrow$E$\rightarrow$H$\rightarrow$L is far from the targets, it should not be prioritized. The efficacy of directed fuzzing is affected when there is more than a single target, as finding the global shortest distance has discrepancy. 

The reason behind such discrepancy is that the distance-based seed measurement only focuses on the shortest path. When there are multiple paths reaching the same target, the longer ones might be ignored, causing discrepancy in the result. In Fig.~\ref{globalbias}, if the paths A$\rightarrow$C$\rightarrow$K and A$\rightarrow$C$\rightarrow$E$\rightarrow$H$\rightarrow$O are considered, then $d_{ACK} = (4/3 + 3/4 + 0)/3 \approx 0.69$, $d_{ACEHO} = (4/3 + 3/4 + 2 + 1 + 0)/5 \approx 1.02$. As expected, $d_{ACK} < d_{ACEHO} < d_{ACEIMNO}$. This is because path A$\rightarrow$C$\rightarrow$K and path A$\rightarrow$C$\rightarrow$E$\rightarrow$H$\rightarrow$O are the shortest paths from A to targets K and O, respectively. The shortest path is always prioritized. 
Such discrepancy is realistic and frequently occurs when three conditions are all met: (1) Multiple targets are measured by distance; (2) At least one target has more than one viable path; (3) A seed exercises the longer path and is measured by this distance. Multi-targets testing is a frequently used scenario when applying DGF. For example, testing patches by setting code changes as targets. Thus, condition 1) is easy to meet. For condition 2), we also use the error handling code as an example. The error-handling code can be the destination of many functional modules, which means a target in the error-handling code is usually reachable via many paths, thus, condition 2) is also easy to meet. Finally, the satisfaction of condition 3) is uncertain as we cannot guarantee the longer path is exercised. Only when a seed exercises the longer path, it would be measured by this distance, and a discrepancy occurs.

To avoid such discrepancy, all potential paths to the targets must be accounted for. For example, under a different
context, the distances from the calling function to the immediately called function may not be exactly the same.
To solve this problem, Hawkeye uses ``adjacent-function distance augmentation'' based on a lightweight static analysis \cite{chen2018hawkeye}, which considers the patterns of the (immediate) call relation based on the generated call graph to augment the distance that is defined by immediate calling relation between caller and callee.
Another strategy for coordinating multi-targets is separating the targets. For each seed, only the minimum distance for all targets is selected as the seed distance, and the seeds are prioritized based on this min-distance~\cite{peng20191dvul}. The effect of this is to negate the possibility of biasing into global optimal solutions but at the cost of increasing the time required to hit a given target.

\subsection{Inflexible Coordination of Exploration Phase and Exploitation Phase}

Another challenge of DGF lies in coordinating the exploration-exploitation trade-off. On the one hand, more exploration is necessary to provide adequate information for the exploitation; on the other hand, an overfull exploration would consume many resources and delay the exploitation. It is difficult to determine the boundary between the exploration phase and the exploitation phase to achieve the best performance. 

Most directed greybox fuzzers, such as AFLGo, adopt a fixed splitting of the exploration phase and the exploitation phase. The time budgets are pre-set in the test configuration before testing. 
Such a scheme is preliminary because the separation point is inflexible and relies on the human experience. Since each PUT is different, such fixed splitting is less adaptive. Once the exploration phase gives way to the exploitation phase, there is no going back even if the direction performance is poor due to insufficient paths.

\begin{figure}
\centering
\includegraphics[width=0.5\columnwidth]{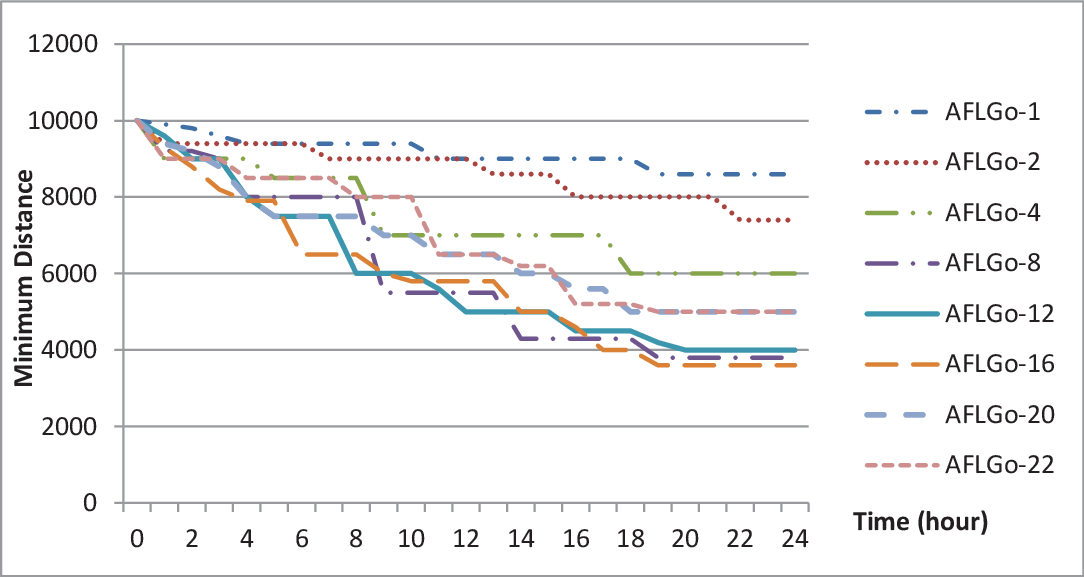}
\caption{Comparison on different splittings of the exploration phase and the exploitation phase.}
\label{ee}
\end{figure}

The efficacy of DGF is determined by how the resources for exploration and exploitation are divided. To elucidate this with a case study, AFLGo was applied to \texttt{libxml} using the ``-z'' parameter of AFLGo to set different time budgets for the exploration phase and compare the performance. As Fig.~\ref{ee} shows, the horizontal coordinate shows the time duration of the test, and the vertical coordinate means the minimum distance of all the generated inputs to the target code areas (min-distance). A small “min-distance” indicates a better-directed performance. The experiments last for 24 hours, and AFLGo-1 means 1 hour of exploration with 23 hours of exploitation, and the rest are similar. From the results, it can be concluded that the splitting of the exploration and exploitation phases affects the performance of DGF, and that the best performance (AFLGo-16) requires adequate time for both of the two phases. However, it is difficult to get optimum splitting.

Among the directed fuzzers investigated, only one work tries to optimize the coordination of exploration-exploitation. RDFuzz~\cite{ye2020rdfuzz} combines distance and frequency to evaluate the inputs. Low-frequency inputs are required in the exploration phase to improve the coverage, while short-distance inputs are favored in the exploitation phase to achieve the target code areas. Finally, an inter-twined testing schedule is used to conduct the exploration and exploitation alternately. However, the classification of the four input types (short/long distance and low/high frequency) is preliminary, and the performance heavily depends on the empirical threshold values.

\subsection{Dependence on the PUT Source Code}
\label{binary}

Most of the known DGF works~\cite{bohme2017directed, chen2018hawkeye, ye2020rdfuzz} are implemented on top of AFL and inherit AFL's compile-time instrumentation scheme to feedback the execution status or calculate distance-based metrics. A significant drawback of such a scheme is the dependence on the PUT source code. Thus, it is unsuitable for testing scenarios where the source code is unavailable, such as commercial off-the-shelf (COTS) software, or the security-critical programs that rely partly on third-party libraries.

There are multiple reasons that hinder the application of DGF at the binary level.
%occasions whereby binary-level DGF is not the best tool for use. 
First, \textbf{heavy run-time overhead}. A straightforward solution to binary-level testing is through a full-system emulator, such as QEMU~\cite{nguyen2020binary}. However, emulator-based tools are usually less efficient. For example, the execution speed of vanilla AFL is 2–5 times faster than its QEMU mode \cite{chen2019ptrix}. 
Second, \textbf{difficulty in collecting target information}. An open-source PUT can be used to obtain target information from various channels, such as the CVE vulnerability descriptions~\cite{wangnot, kim2019poster}, changes made in the git commit logs \cite{you2017semfuzz}, and human experience on critical sites in the source code. 
%However, for binary code, only target information from bug traces can be extracted \cite{nguyen2020binary}. 
However, for binary code, we can only extract target information from bug traces \cite{nguyen2020binary}.
Third, \textbf{difficulty in labeling the targets}. For the source code instrumentation approach, the targets can be labeled based on the source code (e.g., cxxfilt.c, line 100). However, it is much more difficult for the binary. Since the binary code is hard to read, it must be disassembled using tools such as IDA Pro~\cite{nguyen2020binary}, and the targets labeled with virtual addresses, which is both inconvenient and time-consuming.

A viable solution to alleviate the performance limitation is hardware assistance, such as Intel Processor Trace (PT). % or Intel Last Branch Record
Intel PT is a lightweight hardware feature in Intel processors. It can trace program execution on the fly with negligible overhead (averagely 4.3x faster than QEMU-AFL~\cite{zhang2018ptfuzz}), which replaces the need for dynamic instrumentation. Using the packet trace captured by Intel PT along with the corresponding binary of the PUT, the execution path of the PUT could be fully reconstructed.
%Previous hardware features such as Intel Last Branch Record also perform program tracing, but its output is stored in special registers instead of the main memory, which limits the trace size. 
There have been attempts of fuzzing with PT~\cite{schumilo2017kafl,zhang2018ptfuzz,chen2019ptrix,swiecki2016honggfuzz}, but it has never been used to DGF yet. 
For the problem of target identification and labeling at the binary code level, machine-learning-based approach \cite{li2019v, zhao2019suzzer} and heuristic binary diffing approach \cite{chen2019savior} can be leveraged to automatically identify the vulnerable code.

\section{Application of Directed Greybox Fuzzing}

DGF has a good application prospect. 
When a practitioner chooses a directed greybox fuzzer, the first thing to consider is the application scenario. We summarize the following typical scenario for the DGF application. 

\textbf{Patch testing}. DGF can test whether a patch is complete and compatible. A patch is incomplete when a bug can be triggered by multiple inputs~\cite{wang2019avpredictor}, but the patch only fixes a part of them. For example, CVE-2017-15939 is caused by an incomplete fix for CVE-2017-15023~\cite{chen2018hawkeye}. Meanwhile, a patch can introduce new bugs~\cite{wang2017double}. For example, CVE-2016-5728 is introduced by a careless code update. Thus, directed fuzzing towards problematic changes or patches has a higher chance of exposing bugs. 
For example, DeltaFuzz \cite{zhangdeltafuzz} and AFLChurn \cite{zhu2021regression} are designed for regression testing.
SemFuzz~\cite{you2017semfuzz} uses code changes from git commit logs. UAFuzz \cite{nguyen2020binary} and 1dvul \cite{peng20191dvul} uses binary-level comparison to identify patch-related target branches, which are particularly suitable for this scenario. 

\textbf{Bug reproduction}. DGF can reproduce a known bug when the buggy input is unavailable. For example, due to concerns such as privacy, some applications (e.g., video player) are not allowed to send the input file. With DGF, the in-house test team can use DGF to reproduce the crash with the method calls in stack-trace and some environmental parameters~\cite{bohme2017directed}. DGF is also helpful when generating Proof-of-Concept (PoC) inputs of disclosed vulnerabilities with given bug report information~\cite{peng20191dvul, you2017semfuzz}. In fact, DGF is in demand because 45.1\% of the usual bug reports cannot be reproduced due to missing information and user privacy violations \cite{mu2018understanding}. TortoiseFuzz~\cite{wangnot} and DrillerGo~\cite{kim2019poster} utilize CVE vulnerability descriptions as target information, while UAFuzz~\cite{nguyen2020binary} extracts target information from bug traces, both of which are suitable for this scenario. 

\textbf{Knowledge integration}. DGF can boost program testing by integrating the knowledge from a human analyst. Human-in-the-loop can help to overcome roadblocks and explore the program’s state space more thoroughly. For example,  IJON~\cite{aschermannijon} use human experience to identify the security-sensitive program sites (e.g., call site of \texttt{malloc()} and \texttt{strcpy()}) to guide fuzzing towards error-prone parts \cite{aschermannijon}, which are suitable for this scenario.

\textbf{Result validation}. DGF can validate the result of other software testing approaches. Testing approaches such as static analysis and machine learning can help to identify potentially vulnerable targets, though the results are inaccurate. DGF can be used to refine the results by removing false positives. Tools like V-Fuzz \cite{li2019v}, SUZZER \cite{zhao2019suzzer}, DeFuzz \cite{zhu2020defuzz}, ParmeSan \cite{osterlundparmesan} are suitable for this scenario.

\textbf{Energy saving}. Another interesting application of DGF is when the testing resource is limited. For example, IoT devices fuzzing. Under this circumstance, identifying critical code areas to guide testing is more efficient than testing the whole program in an undirected manner, which can save time and computational resources being spent on non-buggy code regions. GREYHOUND \cite{garbelini2020greyhound} and RVFUZZER \cite{kim2019rvfuzzer} are designed for Wi-Fi client and robotic vehicles respectively, and are both suitable for this scenario.

\textbf{Non-crash bug detection}. Finally, DGF can detect non-crash bugs based on customized indicators. For example, it can find uncontrolled memory consumption bugs under the guidance of memory usage \cite{wen2020memlock}, and find algorithmic complexity vulnerabilities under the guidance of resource usage \cite{petsios2017slowfuzz,liunderstanding}.
%GREYHOUND \cite{garbelini2020greyhound} tests anomalous behaviors of Wi-Fi client and RVFUZZER finds input validation bugs in robotic vehicles,

The second thing to consider is the test conditions. Of these, the source code availability is of vital importance.
In order to realize directed fuzzing, researchers use additional instrumentation and data analysis in the fuzzing process. Taking AFLGo as an example, when instrumenting the source code at compile-time, the control-flow and call graphs are constructed via LLVM's link-time-optimization pass. After this, AFLGo measures the distance between each basic block and a target location by parsing the call graph and intra-procedure control-flow graph of the PUT. For the tools reviewed herein, 81\% rely on the PUT source code.

Since both parsing graphs and calculating distances are very time-consuming, pre-processing is required. AFLGo moves most of the program analysis to the instrumentation phase at compile-time in exchange for efficiency at run-time. Notwithstanding this, AFLGo still spent nearly 2 hours compiling and instrumenting \texttt{cxxfilt} (Binutils)~\cite{nguyen2020binary}, which is a non-negligible time cost. For cases where the source code is unavailable, there are three challenges to consider—the heavy run-time overhead caused by QEMU~\cite{nguyen2020binary}, the difficulty in collecting target information, and the difficulty in labeling targets, all of which result in inconvenience and reduced efficiency (this is discussed in detail in Section \ref{binary}).

Last but not least, the number of targets and the number of testing objectives also affect the choice of a tool. When there are multiple targets, the relationship among targets is also exploitable. For example, UAFL~\cite{wang2020uafl} takes into account the operation ordering of target sequences to find complex behavioral use-after-free vulnerabilities (will discuss in Section \ref{relation}).
Most of the tools investigated tend to only focus on optimizing a single objective, such as covering specific targets. Multi-objective optimization is a practical option that meets the demand of optimizing more than one fitness metric simultaneously. For example, Memlock \cite{wen2020memlock} generates test sets that maximize memory usage and code coverage at the same time (will discuss in detail in Section \ref{object}).

\section{Threats to Validity}
First, our search method in this survey mainly focuses on the top venue works, which might miss some works that are related to DGF but not published in top venues. Second, since this survey was finished by 2022.5, it only collected papers published from 2017.1 to 2022.5, thus, it might miss some new works that were published after 2022.5. These new works might bring new techniques that can address the challenges proposed in this paper. Nevertheless, this paper can still reflect the main research progress and future trends. 

\section{Related Works}
\textbf{Fuzzing surveys}. 
So far there have been several surveys on fuzzing. As far as we can find, Fell \cite{fell2017review} conducted the first review on fuzzing in 2017, which introduced the basic scheme of fuzzing and the existing tools on protocols fuzzing and web browser fuzzing. 
Li \textit{et al.} presented an overview of fuzzing solutions by the year 2017 and discussed techniques that could make the fuzzing process smarter and more efficient, including static analysis, taint analysis, machine learning, and format method. 
Liang \textit{et al.} \cite{liang2018fuzzing} summarized 18 typical fuzzers ranging from 2004 to 2017. They discussed the key obstacles and some state-of-the-art technologies that aim to overcome or mitigate these obstacles. 
In 2019, Man{\`e}s \textit{et al.} \cite{manes2019art} gave a detailed survey on 63 modern fuzzers. They explored the design decisions at every stage of fuzzing by surveying the related literature.
B{\"o}hme \textit{et al.} \cite{bohme2020fuzzing} summarized the open challenges and opportunities for fuzzing and symbolic execution as they emerged in discussions
among researchers and practitioners in a Shonan Meeting.
Finally, Zhu  \textit{et al.} \cite{bohme2020fuzzing} gave the most up-to-date survey on fuzzing to narrow down the gaps between the entire input space and the defect space.  The survey reviews and analyzes the gaps as well as their solutions, considering both breadth and depth.
In addition to surveys on common fuzzing, there are also surveys focused on subclasses of fuzzing. 
Eisele \textit{et al.} reviewed the fuzzing approaches for embedded systems \cite{eisele2022embedded}. They gave a formal definition of embedded fuzzing and grouped the approaches according to how the execution environment is served to the system under test.
Saavedra  \textit{et al.} \cite{saavedra2019review} reviewed the machine learning applications in fuzzing, including deep learning, neural networks, and reinforcement learning. They discussed successful applications of machine learning to fuzzing, such as input generation, seed selection, and corpus minimization. 
Wang  \textit{et al.} also gave a review of fuzzing based on machine learning techniques \cite{wang2020systematic}. They identified six different stages in which machine learning has been used and studied the machine learning-based fuzzing models from the selection of machine learning algorithm, pre-processing methods,
datasets, evaluation metrics, and hyperparameters setting. They also assessed the performance of the machine learning models based on the frequently used evaluation metrics.
Zhang \textit{et al.} also gave a preliminary survey on directed fuzzy technology \cite{zhang2018survey}. However, this survey only gave a brief introduction and lacked a detailed comparison of different approaches. 

\textbf{Generational fuzzing}.
Generating syntactically and semantically valid inputs can improve the efficiency of fuzzers in code coverage and bug detection \cite{nguyen2022bedivfuzz}. To achieve this goal, researchers have proposed generation-based fuzzing, which utilizes grammar or models to describe the input structure and generate syntactically correct inputs \cite{godefroid2008grammar,nguyen2022bedivfuzz,padhye2019semantic, peach,holler2012fuzzing}. This approach has been widely employed in fuzzing targets that require highly structured inputs, such as parsers, protocols, and compilers \cite{aschermann2019nautilus,jero2019leveraging,godefroid2008grammar,nguyen2022bedivfuzz, peach,holler2012fuzzing}. Utilizing a well-defined grammar or model can significantly enhance the fuzzer. However, creating a manual grammar requires substantial effort \cite{xu2020freedom,park2020fuzzing,domfuzzer}. To alleviate this burden, some researchers have explored learning the grammar from existing test cases using machine learning techniques \cite{wang2017skyfire,liu2019deepfuzz,godefroid2017learn}. Nevertheless, as pointed out in \cite{nguyen2022bedivfuzz}, having only syntactically correct inputs may not be sufficient to explore deeper regions of programs. Recent studies have combined mutation-based fuzzing with generation-based fuzzing and incorporated coverage mechanisms in greybox fuzzing to improve the efficiency of fuzzers. Zest \cite{padhye2019semantic} integrates coverage feedback to generate inputs with high semantic coverage. BeDivFuzz \cite{nguyen2022bedivfuzz} follows a similar mechanism but extends it with structural mutation strategies. DGF can also be enhanced by incorporating generation-based techniques to generate highly structured test cases that satisfy more constraints and reach the intended targets.

\textbf{Anti-fuzzing}.
Whitehouse \textit{et al.} \cite{whitehouse2014introduction} introduced the concept of anti-fuzzing and proposed strategies such as fake crashes and performance degradation. David et al. \cite{edholm2016escaping} identified four attack vectors against fuzzers, including execution speed, crash masking, fuzzer detection, and feedback mechanism detection. FUZZIFICATION \cite{jung2019fuzzification} and ANTIFUZZ \cite{guler2019antifuzz} subsequently proposed countermeasures for degrading fuzzers. For example, they injected a large number of fake bugs into the target application to attack the feedback mechanism in CGF, transformed explicit data flows into implicit data flows for anti-hybrid fuzzing, and inserted delay codes in cold paths to slow down the fuzzer. Some of these countermeasures are also employed by VALL-NUT \cite{li2021vall}. However, certain strategies, such as executing delay codes, are only applied when the inputs trigger paths that regular users rarely reach but fuzzers are prone to fall into. In the case of DGF, since the scheduling process of target-directed greybox fuzzing is guided by predefined targets, a target that is not present in those paths may result in the fuzzer generating fewer test cases that cover the cold paths. As for behavior-directed greybox fuzzing, such as memlock, memory consumption is the fitness metric, which is barely affected by anti-fuzzing techniques. This may mitigate the effectiveness of these path-based strategies in anti-fuzzing, and these passive anti-fuzzing strategies may have a lesser impact on DGF compared to CGF.
Nevertheless, No-Fuzz \cite{zhou2022no} proposed strategies for accurately detecting binary-only instrumentations used by fuzzers, such as timing-related techniques and execution frequency examination. Once fuzzers are detected, mitigation techniques are implemented. Jiang et al. \cite{jiang2022examiner} utilized inconsistent instructions to detect the emulation technique used in binary-only fuzzing. These techniques have been proven to have strong detection capabilities for fuzzers and anti-fuzzing, including DGF.

\textbf{Fuzzing cost}.
%In recent years, two state-of-the-art techniques, namely BEACON \cite{huangbeacon} and SelectFuzz \cite{luo2023selectfuzz}, have significantly improved the speed of bug exposure by reducing the cost of fuzzing. 
BEACON \cite{huangbeacon} leverages symbolic execution to analyze the feasibility of different paths and eliminates those that cannot lead to the target, thereby reducing the overall fuzzing cost. In a similar vein, SelectFuzz \cite{luo2023selectfuzz} conducts a preliminary analysis of the reachability of basic blocks and selectively instruments and calculates seed distances for blocks that are reachable. As a result, the overhead associated with instrumentation and seed distance calculation is minimized.  %By this means, SelectFuzz avoids exploring irrelevant code, further reducing the cost of fuzzing. By adopting these strategies, both BEACON and SelectFuzz effectively decrease the fuzzing cost by minimizing the exploration of code that does not contribute to reaching the desired targets, thereby enhancing the speed of bug exposure.
In contrast, other techniques such as Windranger, CAFL, and FuzzGuard actually increase the cost of fuzzing due to their requirements for analyzing and collecting DBBs and path constraints or collecting and filtering seeds. However, since the fitness metrics and fuzzing strategies proposed by these techniques can effectively guide DGF to reach targets faster, considering the improved speed of bug exposure achieved by these fitness metrics, the additional cost of fuzzing is deemed acceptable, which can still achieve a tradeoff between fuzzing cost and bug exposure.

\section{Conclusion and Perspectives on Future Trends}
%This section provides an overview of the prevailing trends in DGF studies. 

Directed greybox fuzzing is a practical and scalable approach to software testing, which can be applied to specific scenarios, such as patch testing, bug reproduction, and special bug detection. Modern DGF has evolved from reaching target locations to detecting complex deep behavioral bugs. This paper conducts the first in-depth study of DGF based on the review of 42 state-of-the-art tools related to DGF. After summarizing the recent progress in DGF and the challenges faced by DGF, we make the following suggestions in terms of the perspectives and future trends of DGF, aiming to facilitate and boost research in this field.

%Based on the above discussion, we recommend taking into account the relationship among targets when selecting and prioritizing targets. The targets with higher reachability should have higher priority. Targets with a closer relationship should be covered with fewer test runs.

\subsection{Exploitation of Relationship between Targets}

\label{relation}
When there are multiple targets in a targeted fuzzing task, how to coordinate these targets is another challenge.
Although 86\% (36/42) of the fuzzers we investigated support multi-targets, only four of them paid attention to the relationship among targets. 
For multiple targets to be reached, exploiting the relationship among targets is meaningful for optimizing DGF. If the targets are unrelated, weights can be assigned to them to differentiate the importance or probability. Alternatively, the hidden relationship can be extracted and exploited to improve directedness. For example, UAFL~\cite{wang2020uafl} takes into account the operation sequence ordering when leveraging target sequences to find use-after-free vulnerabilities. This is because, to trigger such behavioral complex vulnerabilities, one needs not only to cover individual edges but also to traverse some longer sequences of edges in a particular order. Such a method can be extended to detect semantic bugs, such as double-free and API misuse. Berry \cite{liang2020sequence} enhanced the target sequences with execution context (i.e. necessary nodes required to reach the nodes in the target sequences) for all paths. Similarly, KCFuzz \cite{wang2021kcfuzz} regards the parent nodes in the path to the target as keypoints to cover.
CAFL \cite{lee2021constraint} regards the data conditions along the path to the target as constraints and drives the seeds to satisfy the constraints in orders to finally reach the target.

\begin{figure}[t]
\centering
\includegraphics[width=0.4\columnwidth]{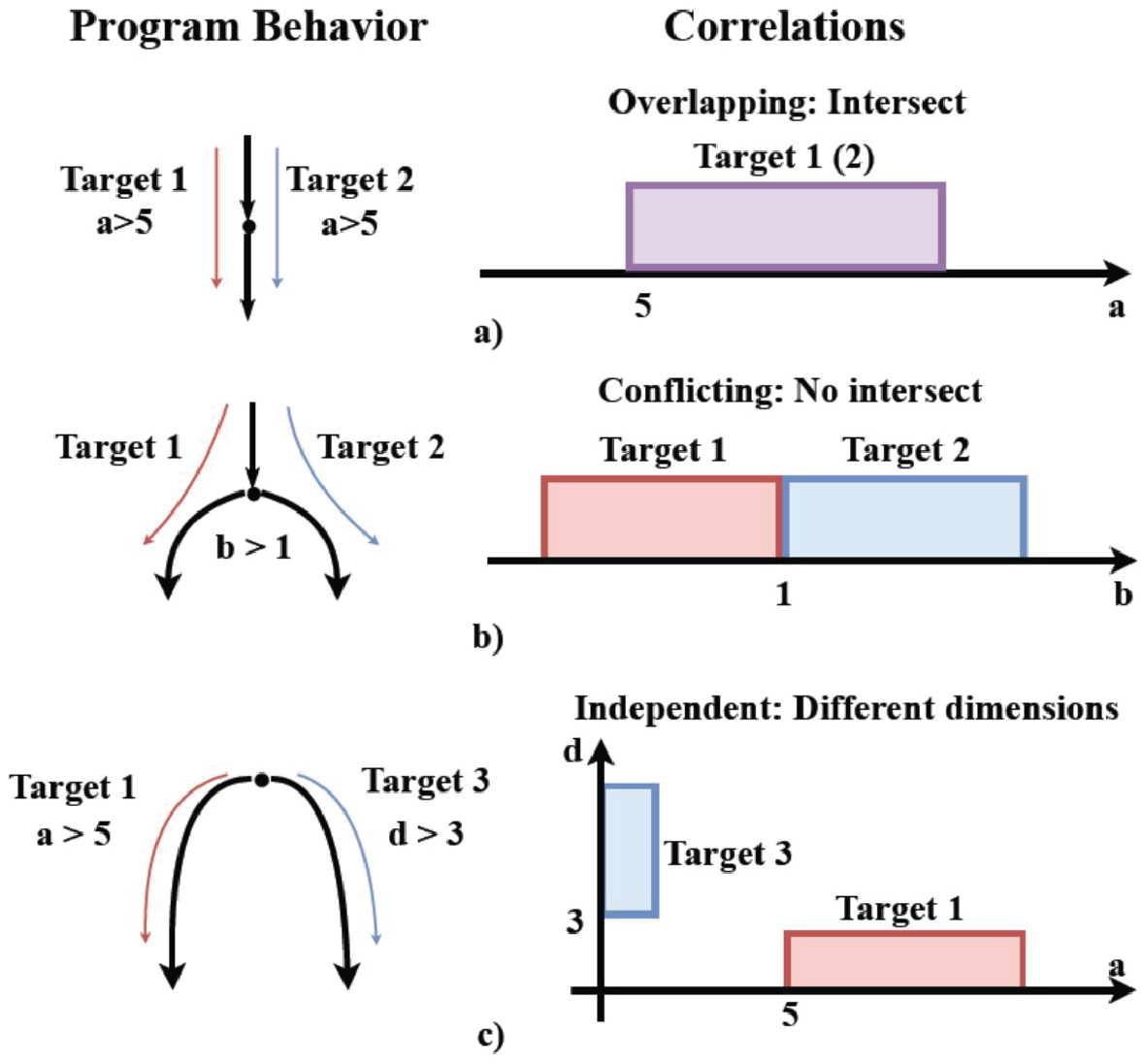}
\caption{Correlations among different targets.}
\label{multitargets}
\end{figure}

\begin{table}[t]
\centering
\begin{tabular}{p{1.2cm}@{}p{10cm}}
%\begin{tabular}{ll}
\hline
 
&Listing 1: A running example for multi-targets. \\   
\hline
01 & \textbf{int} foo()\{\\
02 & \quad \textbf{signed} a,b,c,d = parse\_input();\\
03 & \quad \textbf{int} flag;\\
04 & \\
05 & \quad \textbf{if}(a>5)\{\\
06 & \quad\quad \textbf{if}(b>1)\{\\
07 & \quad\quad\quad flag=0;\\
08 & \quad\quad \}\\
09 & \quad\quad \textbf{else} \{\\
10 & \quad\quad\quad flag=1;\\
11 & \quad\quad\quad \textbf{if}(...) \{...\} //target-irrelevant\\
12 & \quad\quad \}\\
13 & \quad\quad ...\\
14 & \quad\quad \textbf{if}(flag)\{\\
15 & \quad\quad\quad target1();\\
16 & \quad\quad \}\\
17 & \quad\quad \textbf{else if}(c>2)\{ //contradict to flag==1\\
18 & \quad\quad\quad target2();\\
19 & \quad\quad \}\\
20 & \quad \}\\
21 & \\
22 & \quad \textbf{if}(d>3)\{\\
23 & \quad\quad target3();\\
24 & \quad \}\\
25 & \}\\
\hline
\end{tabular}
\label{list1}
\end{table}

Huang \textit{et al.} proposed to exploit the correlations between different targets \cite{huangtitan}, primarily in the form of path condition
\textit{overlapping}, \textit{conflicting}, and \textit{independence}, as illustrated in Figure \ref{multitargets}.
We use their running example in Listing 1 to show the three correlations \cite{huangtitan}.
(1) Condition a > 5 is an overlapping condition for both targets 1 and 2
because reaching them both requires the condition to hold.
The overlapping condition can direct the fuzzer to cover the true branch at Line 5.
(2) Reaching targets 1 and 2 have mutually exclusive demands
on the condition of b, namely b $\leqslant$ 1 and b >1, which we
refer to as a conflicting condition for the two targets.
We regard the seeds satisfying b $\leqslant$ 1 as more
likely to cover target 1 and seeds satisfying b > 1 are more likely to cover target 2.
The fuzzer can more accurately select seeds for multiple
targets by using conflict correlations to differentiate the 
difficulties in reaching different targets, e.g., targets 1 and 2. 
(3) The condition d > 3 at Line 22 only influences the reachability
of target 3 but does not affect whether targets 1 and 2
are reached. Similarly, condition a >5 affects reaching
targets 1 and 2 but not target 3. As a result, we can
mutate the independent bytes simultaneously, which could
help approach multiple targets with fewer executions. 
%If we simultaneously mutate d with other bytes, such as a for
%targets 1 and 2, the fuzzer can attempt to approach all three
%targets in a single execution because the conditions for d
%and a do not influence each other.

Herein, we suggest that the following relationships can be considered for DGF research.

\textbf{The spatial relationship}. Namely the relative position of targets on the execution tree. Consider the relation between two targets, including whether they occupy the same branch, the level of shared executions, and their relative precedence if any. 

\textbf{The state relationship}. For targets that involve the program state, consider their position in the state space. For example, whether two targets share the same state, and whether two states can convert to each other on the state transition map.

\textbf{The interleaving relationship}. For multi-threaded programs, thread scheduling also affects the execution ordering of events in different threads. Targets that can be reached under the same thread interleaving should have a close relationship in the interleaving space.

\subsection{Design Multi-dimensional Fitness Metric}
Current fuzzing approaches mainly focus on the coverage at the path level, such as maximizing the overall path coverage or reaching specific code parts, which neglects the fact that some bugs will not be triggered or manifest even when vulnerable code is exercised. For example, a buffer overflow vulnerability will be exhibited at a buffer access location only when the buffer access pointer points outside the buffer. Similarly, an integer overflow vulnerability will be observed at a program location only when the variable being incremented has a large enough value. To detect such ``hard-to-manifest'' vulnerability, the fitness metric must be extended to be multi-dimensional, such as the state space.

%\subsubsection{State Space}
In practice, exploring a complex state machine is difficult, and most fuzzing-based approaches only make progress when exercising certain code, neglecting the update of the state machine and would not fuzz the corresponding test input further. %There is potential for certain vulnerabilities to be inconsistently revealed, whereby only certain executions that reach the vulnerability point with the right state may show the vulnerable behavior. 
However, some vulnerabilities may not get revealed for every visit to the program point. Only certain executions that reach the vulnerability point with the right state may exhibit the vulnerable behavior.
To expose such vulnerability, we need inputs that not only reach the vulnerability location but also match the vulnerable state~\cite{medicherla2020fitness}. 

%\begin{equation*}\label{headr}
\begin{equation}
\label{headr}
\begin{aligned}
headroom=\begin{cases}
0,&if A_c \geqslant A_h+s;\\
(A_h+s-A_c)/s,&if A_h \leqslant A_c < A_h +s;      \\
1, &otherwise.
\end{cases}
\end{aligned}
\end{equation}

In order to find hard-to-manifest vulnerability (e.g., buffer overflow and integer overflow), AFL-HR~\cite{medicherla2020fitness} defines a fitness metric ranging from 0 to 1, called \textit{headroom}, to indicate how closely a test input can expose a potential vulnerability at a given vulnerability location. 
For example, for buffer overflow vulnerabilities, we consider the buffer access location is $v_l$, where $ptr$ is the pointer, $A_c$ is the value of $ptr$ during the visit, $A_h$ is the starting address of the allocated buffer, and $s$ is the size of the allocated buffer.
As Equation \ref{headr} shows, the headroom is defined as the minimum distance between the location pointed to by the buffer access pointer and the end of the buffer across all visits to this location, divided by the size of the buffer.

IJON~\cite{aschermannijon} leverages an annotation mechanism that allows a human analyst to help overcome roadblocks and explore the program’s state space more thoroughly. %The result is that developers must note the state space dimension while determining whether a given vulnerability can be reached. 
Thus, state space is a dimension that is worth taking into account as a fitness metric alongside the reachability of the vulnerability location.

\subsection{Multi-objective Optimization}
\label{object}
For simplicity, the vast majority of contemporary studies have opted to ignore the possibility of multi-objective targeting through the simultaneous application of a range of metrics. For example, a tester might be interested in achieving higher coverage, but while also targeting unusually long execution times, security properties, memory consumption, or energy consumption. Multi-objective optimization provides an advantage over traditional policies that are only capable of achieving one goal. It formulates the trade-off among multiple properties, such as usability and security \cite{harman2015achievements}. For example, multi-objective optimization can generate test sets that cover specific targets while also maximizing overall coverage, or prioritizing tests that cover as much of the software as possible whilst minimizing the amount of time that tests take to run~\cite{mcminn2011search}. The result of a multi-objective search is a set of Pareto-optimal solutions, where each member of the set is no better than any of the others for all of the objectives~\cite{mcminn2011search}.

Multi-objective optimization is an open problem in the SBST community~\cite{mcminn2011search}, which also is a challenge for DGF. A general solution of optimizing multiple objectives is co-evolution, where two (or more) populations of test inputs evolve simultaneously in a cooperative manner using their own fitness functions 
For example, AFL-HR~\cite{medicherla2020fitness} defines the fitness metric \textit{headroom} to measure how closely a test input can expose a potential vulnerability at a given vulnerability location. Then it uses a co-evolutionary computation model to evolve test inputs for both coverage-based and headroom-based fitness metrics simultaneously. Similarly, other fitness metrics such as memory consumption~\cite{wen2020memlock} and new maxima of execution counts \cite{lemieux2018perffuzz} have also been applied in a co-evolutionary manner. In contrast, FuzzFactory \cite{padhye2019fuzzfactory} provides a framework that supports multiple domain-specific objectives that are achieved by selecting and saving intermediate inputs from a custom predicate, which avoids the non-trivial implementation of mutation and search heuristics.

To show how multi-objective optimization works. We use MobFuzz \cite{Zhang2022MobFuzz}  as an example.
MobFuzz models the multi-objective optimization as the problem of selecting the best objective combination.
The fuzzing process is divided into $t$ intervals that each lasts for one minute. In the initial phase, the fuzzer selects the objective combinations in order 
and score them by the rewards as Equation \ref{mob} defines \cite{Zhang2022MobFuzz}. 

%\begin{scriptsize}
	\begin{equation}
		\label{mob}
		\begin{aligned}
			Score(C_l, t)&=\overline{R}(C_l, t)+U(C_l, t)\\
			&=\frac{\sum_{k=0}^{t}{R(C_l, k)}}{t}+\gamma*\sqrt{\frac{ln(\sum_{C_l\in{C}}^{}{n_l})}{n_l}}
		\end{aligned}
	\end{equation}
%\end{scriptsize}

In Equation \ref{mob}, MobFuzz applies the UCB1 \cite{agrawal1995sample} algorithm to select the combinations with the highest scores. In UCB1, $Score(C_l, t)$ denotes the final score for the combination to make decisions, which consists of two components: $\overline{R}(C_l, t)$ and $U(C_l, t)$. 
$C$ denotes all the objective combinations, and $C_l$ denotes the $l$th combination.
$\overline{R}(C_l, t)$ is the average reward of $C_l$ in previous $t$ rounds, which gives combinations with greater historical rewards higher scores (exploitation). $R(C_l, k)$ is the reward of $C_l$ in the $k$th round. $U(C_l, t)$ is the upper confidence bound of $C_l$, and it adds greater scores to combinations with smaller $n_l$ values (the number of times the combination is selected), which is exploration. At the beginning of fuzzing, MobFuzz goes through an initial stage, in which each objective combination is selected once. After this stage, the $n_l$ value of each combination will be 1. Next, at the end of each round of fuzzing, MobFuzz calculates the score of each combination and chooses the one with the maximum score as the objective combination for the next round. Moreover, $\gamma$ is an empirical parameter in UCB1 that controls the balance between exploration and exploitation. 

Then, it adaptively selects the objective combination that contains the most appropriate objectives for the current situation, 
the best objective combination that has the highest reward is selected and given more power. 
Finally, an evolutionary algorithm is designed for multi-objective optimization in MobFuzz. 
The basic process is as follows:
First, an initial population of seeds with a scale of $N$ is selected.
Next, the offspring seeds are obtained with crossover and mutation among the initial population. 
Then, execute the target program with each seed in the population and obtain related information.
From the second generation onwards, the parent population and the offspring are combined to perform non-dominated sorting.
The seeds with updated objective values are selected to form a new parent population with a scale of $N$.
Finally, the new offspring seeds are generated by the crossover and mutation among the new parent population.
This process repeats until the predefined number of iterations is met.

\begin{figure}
\centering
\includegraphics[width=0.6\columnwidth]{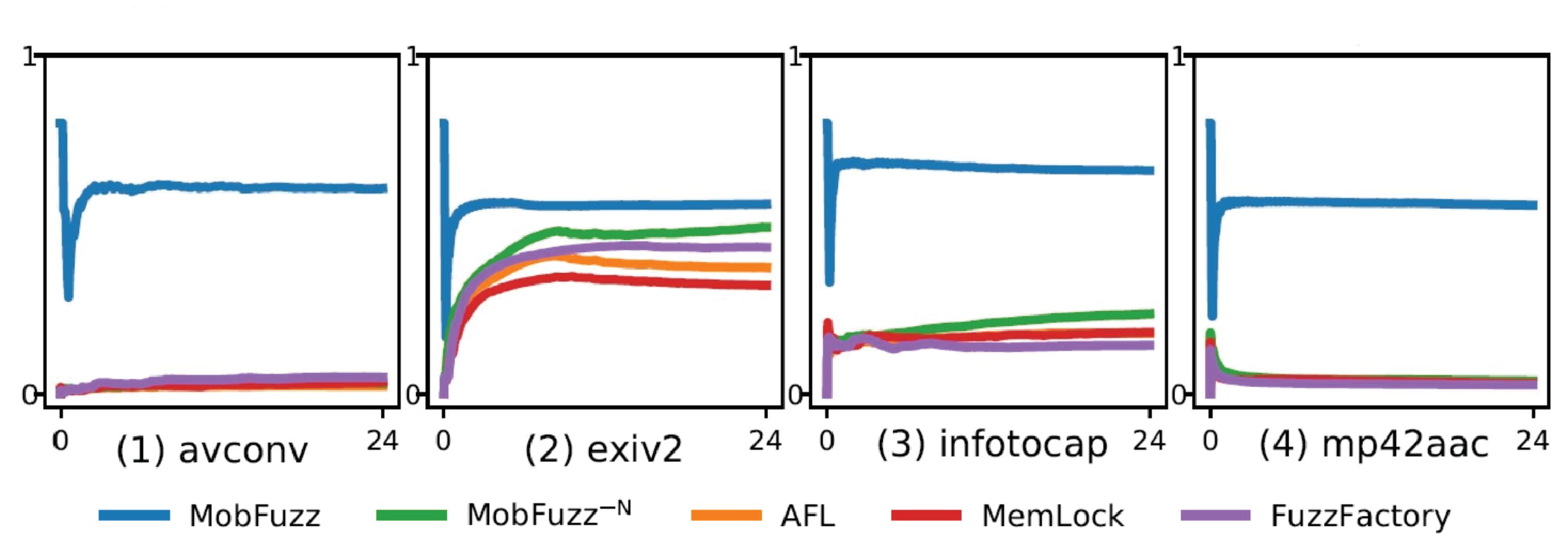}
\caption{Multi-objective optimization performance on good seeds}
\label{goodseeds}
\end{figure}

To show the effectiveness of multi-objective optimization, we use metric \textit{good seeds} to test MobFuzz and baseline fuzzers on four programs.
Seeds that achieve greater objective values than the average in the selected objective combination are defined as \textit{good seeds}. 
In Figure \ref{goodseeds}, the x-axis is the test time, and the y-axis is the percentages of good seeds generated. MobFuzz-N means MobFuzz disables multi-objective optimization. 
From the figure, we can conclude that by multi-objective optimization, the percentages of good seeds in MobFuzz are significantly greater than those of the baseline fuzzers.

\subsection{Target for New Domains}

Among the tools evaluated, only one (SemFuzz \cite{you2017semfuzz}) supports kernel code testing. Thus, introducing DGF to kernel code and guiding fuzzing towards critical sites such as syscalls and error handling codes to find kernel bugs should be a productive direction. Except for kernel testing, protocol testing is also suitable for DGF. Directed testing can strengthen the critical fields of the protocol message, such as the message length and control information.
Zhu \textit{et al} \cite{zhu2021constructing} utilize DGF to construct more complete control flow graphs by targeting and 
exercising indirect jumps.
It is delightful to see that DGF has been applied in the targeted testing for Register Transfer Level (RTL) designs \cite{canakci2021directfuzz}. Hopefully, DGF would be applied to more domains in the future.

Although DGF has been trying to discover new bug types, such as use-after-free and memory consumption bugs, many commonly seen bug types have not yet been included. Thus, another research direction is applying DGF to bug types with specific feature, such as information leakage, time-of-check to time-of-use \cite{wei2005tocttou}, and double-fetch bugs\cite{wang2017double, wang2018survey}. For example, to detect a double-fetch bug, DGF would be useful to guide the testing towards code parts that launch continuous kernel reads of the same user memory address.

\section*{Acknowledgement}
The authors would like to sincerely thank all the reviewers for your time and expertise on this paper. Your insightful comments help us improve this work. This work is partially supported by the National University of Defense Technology Research Project (ZK20-17, ZK20-09), the National Natural Science Foundation China (62272472, U22B2005, 61972412), and the HUNAN Province Natural Science Foundation (2021JJ40692).

%\subsection*{Author contributions}

%This is an author contribution text. This is an author contribution text. This is an author contribution text. This is an author contribution text. This is an author contribution text.

\subsection*{Financial disclosure}

None reported.

\subsection*{Conflict of interest}

The authors declare no potential conflict of interests.

\nocite{*}% Show all bib entries - both cited and uncited; comment this line to view only cited bib entries;
\bibliography{wileyNJD-AMA}

\clearpage

%\section*{Author Biography}

%\begin{biography}{\includegraphics[width=66pt,height=86pt,draft]{empty}}{\textbf{Author Name.} This is sample author biography text this is sample author biography text this is sample author biography text this is sample author biography text this is sample author biography text this is sample author biography text this is sample author biography text this is sample author biography text this is sample author biography text this is sample author biography text this is sample author biography text this is sample author biography text this is sample author biography text this is sample author biography text this is sample author biography text this is sample author biography text this is sample author biography text this is sample author biography text this is sample author biography text this is sample author biography text this is sample author biography text.}
%\end{biography}

\end{document}